\documentclass[twocolumn]{aastex701}
\usepackage{amsmath}
\usepackage{graphicx}	
\usepackage{chemformula}
\usepackage{verbatim}
\usepackage{hyperref,natbib,amsmath,color,ulem}
\usepackage{soul}
\usepackage{color,epstopdf,placeins}
\usepackage{relsize}
\usepackage{subcaption}
\usepackage{ctable}
\usepackage[table]{xcolor}
\usepackage{makecell}
\usepackage{colortbl}
\usepackage{orcidlink}
\usepackage{ulem}
\usepackage[T1]{fontenc}
\usepackage{tabularx}
\graphicspath{{./}{Figures/}}
\usepackage{cancel}
\usepackage{xcolor}

\newcommand{\x}{\sout}

\shorttitle{Dust In Wind-Driven Disks}
\shortauthors{Goksu, Pudritz and Delage}

\begin{document}


\title{Evolution of Dust In Protoplanetary Disks With Magnetized Winds}

\author[orcid=00009-0002-4988-2545,sname=Goksu,gname=Olcay Ates]{Olcay Ates Goksu}
\affiliation{Department of Physics and Astronomy, McMaster University, Hamilton, ON L8S 4M1, Canada}
\email[show]{goksuo@mcmaster.ca}  

\author[orcid=00000-0002-7605-2961,gname=Ralph, sname=Pudritz]{Ralph E. Pudritz} 
\affiliation{Department of Physics and Astronomy, McMaster University, Hamilton, ON L8S 4M1, Canada}
\email{pudritz@mcmaster.ca}

\author[orcid=0000-0002-7901-6157,gname=Timmy,sname=Delage]{Timmy N. Delage}
\altaffiliation{Royal Society Career Development Fellow}
\affiliation{Astrophysics Group, Department of Physics, Imperial College London, Prince Consort Road, London SW7 2AZ, UK}
\email{timmy.delage17@imperial.ac.uk}

\begin{abstract}

Observations and theory increasingly indicate that protoplanetary disks are weakly turbulent, implying that magnetized disk winds may dominate their angular momentum transport and long-term evolution. Yet dust evolution in wind-dominated disks remains poorly understood, despite its importance for planet formation. In particular, whether wind-dominated disk evolution can create regions with enhanced dust-to-gas mass ratios that enable planetesimal formation via the streaming instability is unclear. We build a parametric hybrid disk evolution model that includes turbulent accretion, wind-driven accretion, and wind-driven mass loss to investigate dust evolution in magnetized protoplanetary disks. We employ this model to carry out a parameter study in which the initial total accretion rate, the wind-driven mass loss rate, and the turbulence strength are varied. We find that although wind-dominated disks are more favorable to dust growth and concentration than turbulence-dominated disks, conditions suitable for the streaming instability arise only within the inner few au and only under the most extreme conditions: very weak turbulence ($\alpha_\text{turb} = 10^{-5}$) where magnetized winds drive most of the accretion ($\dot{M}_\text{acc,tot} \simeq \dot{M}_\text{wind,acc}$), and wind-driven mass loss comparable to the accretion rate ($\dot{M}_\text{wind,outflow} \simeq \dot{M}_\text{acc,wind}$). Our results therefore imply that, beyond the inner few au, the parametric wind-driven evolution models considered in the literature do not by themselves produce conditions suitable for planetesimal formation via the streaming instability, motivating more physically complete models in which spontaneous, long-lived dust traps can arise.

\end{abstract}

\keywords{\uat{Stellar accretion disks}{1579} --- \uat{Exoplanet formation}{492} --- \uat{Magnetohydrodynamics}{1964} --- \uat{Circumstellar dust}{236} --- \uat{Dust physics}{2229}}


\section{Introduction} \label{sec:intro}

Protoplanetary disks (PPDs) set the initial conditions for planetary systems, via the interaction between their gas and dust reservoirs during the disk's lifetime. Assuming that no significant dust growth occurs during cloud collapse \citep[which is currently a matter of debate, see ][]{kwon09,silsbee22,lombart26}, young PPDs inherit the dust properties provided by the interstellar medium (ISM), with characteristic sizes of $\sim 1 \ \mu \mathrm{m}$ \citep{mathis77} and an average dust-to-gas mass ratio of $\sim 0.01$ if inherited from the Milky Way ISM \citep{bohlin78}. ISM-like solid dust particles in PPDs require highly efficient growth to reach kilometer-sized planetesimals, from which the subsequent steps of planet formation may proceed via pebble and planetesimal accretion as well as gas accretion \citep{lau22}. However, dust growth is hampered by barriers such as collisional fragmentation \citep{blum2000,birnstiel10} and radial drift \citep{weidenschilling77,birnstiel09}. A potential way to circumvent these barriers is through the streaming instability \citep[SI,][]{youdin2007,johansen09,bai10} -- one of the resonant drag instabilities \citep{squire18} -- in which the rapid concentration of dust into filaments can lead to gravitational collapse of clumps within them into planetesimals.

Whether the SI can operate depends sensitively on both the dust-to-gas mass ratio and the aerodynamic coupling of dust to the gas, which is expressed through the Stokes number \citep{li21,rucska23,lim24}. Larger dust particles can trigger the SI at lower dust-to-gas mass ratios, whereas smaller particles require stronger concentration. Achieving these conditions is challenging because: (1) dust particles with Stokes numbers of order unity undergo rapid radial drift before large dust enhancements can be achieved \citep{brauer08,birnstiel09}; (2) Smaller dust particles remain more tightly coupled to the gas and are therefore more sensitive to turbulent diffusion, which opposes both radial and vertical concentrations \citep{armitage20}.

Gas dynamics in PPDs has traditionally been modeled in terms of turbulent angular momentum transport \citep{shakura73}, driven by hydrodynamic (HD) processes \citep{lesur23,pfeil21,lyra14} and magnetohydrodynamic (MHD) processes such as the magnetorotational instability \citep[MRI;][]{lesur23,balbus91,delage22,delage23}. When interpreted within such turbulent disk models, disk lifetimes \citep{hernandez07} and stellar accretion rates \citep{hartmann98} implied effective transport parameters of the order $\alpha_\mathrm{eff} \sim 10^{-4} - 10^{-2}$. However, new observational methods now allow for  turbulence strengths to be inferred in resolved disks by line broadening \citep{flaherty18}, dust ring widths \citep{dullemond18}, and vertical dust settling \citep{villenave25}. These ALMA observations suggest weak turbulence strengths in the range $\alpha_\mathrm{turb} \sim 10^{-4} - 10^{-3}$. Theoretically, although the MRI can generate strong turbulence in the ideal MHD regime with $\alpha_\mathrm{turb} > 10^{-2}$ \citep{salvesen16}, non-ideal MHD effects such as Ohmic resistivity and ambipolar diffusion will be present in poorly ionized PPDs. These suppress vigorous MRI-driven turbulence over large regions of PPDs \citep{bai13,suriano17,bethune17,delage22,delage23,aoyama23}. Hydrodynamic instabilities such as vertical shear instability \citep{goldreich67,nelson13} and convective overstability \citep{klahr14,lyra14} also tend to generate relatively weak turbulence, with $\alpha_\mathrm{turb}$ in the range of $10^{-5}-10^{-3}$ \citep{lyra14,flock17}. 

These developments suggest that turbulent stresses alone cannot account for the observed and inferred accretion rates of PPDs. Magnetized disk winds, long known to influence disk evolution \citep{blandford82,pudritz86,pelletier92}, have therefore received renewed attention as the potential dominant mechanism for angular momentum removal and gas accretion \citep{bai13,bethune17,suriano17}. This paradigm shift has motivated parametric prescriptions for wind-driven accretion and mass loss that can be implemented in computationally efficient one-dimensional disk evolution models \citep{suzuki16,chambers19,tabone22}, allowing the secular evolution of wind-dominated disks to be explored over millions of years that is not yet accessible to 3D global nonideal MHD models.

MHD wind-dominated weakly turbulent PPDs may be especially important for dust evolution. Weak turbulence reduces radial and vertical dust diffusion, allowing dust to settle and concentrate more efficiently \citep{dubrulle95,youdin07}. MHD wind-driven accretion and mass loss can also alter the gas surface density, which can reduce radial drift, increase dust retention, and increase dust-to-gas mass ratios. For example, in the case of extreme mass loss rates exceeding the wind-driven accretion rate, inner cavities produced by outflows can generate pressure maxima in the inner few AU \citep{suzuki16,weder23}, where dust can accumulate and potentially form planetesimals via the SI \citep{ueda21}. In addition, gas surface accretion due to magnetized disk winds, combined with dust settling onto the midplane, can allow disks to retain dust while the gas is accreted onto the central star, increasing dust-to-gas mass ratios \citep{okuzumi25}. 

A number of studies have begun to discern the dust response to wind-dominated disk evolution in the past decade. Multi-dimensional MHD simulations which include dust evolution have revealed MHD winds' positive effects on dust concentration, a pre-requisite for planetesimal formation through the SI channel \citep{riols20,hsu25}. While highly relevant for uncovering the fine details of MHD-driven gas dynamics and their knock-on effects on dust behavior, such multi-dimensional simulations are limited by computational cost, and thus cannot rigorously model collisional dust growth on timescales comparable to the disk lifetime. One-dimensional simulations can resolve these problems by sacrificing on the finer details of disk dynamics, allowing for broader studies on planet formation and evolution. A number of such studies already exist in the literature \citep{taki21,kawasaki25,weder26}, to which we add the advantages of our model in its ability to explore a broad range of turbulence strengths, wind-driven accretion and outflow strengths, and total accretion rates over Myr-timescales in order to provide a more complete picture accounting for the diversity of observed and simulated PPDs.

In this work, we investigate whether PPDs with magnetized winds can generate regions of both enhanced dust-to-gas mass ratios and suitable particle sizes in which planetesimal formation via the SI could occur. We build a parametric hybrid disk evolution model that includes turbulent accretion, wind-driven accretion, and wind-driven mass loss. For that, we use the one-dimensional disk evolution code \textit{DustPy} \citep{stammler22} to model dust coagulation/fragmentation and dust transport, and modify the gas evolution to include the effect of magnetized disk winds following the formalism of \citet{chambers19}. We employ it to model the evolution of the disk on a timescale of $1\,$Myr, and carry out a parameter study in which we vary the initial total accretion rate over $\dot{M}_\text{acc,init} = 10^{-7}-10^{-9} \ M_\odot/\text{yr}$ in accordance with the observations \citep{manara23}, the MHD wind-driven mass loss rate over $K = 0-0.6$ (with the two ends of the range corresponding to negligible mass loss from the wind and heavy winds at $\dot{M}_\text{wind,outflow} \sim \dot{M}_\text{wind,accretion}$), and the turbulence strength over $\alpha_\text{turb} = 10^{-5}-10^{-3}$ as constrained by recent observations \citep{flaherty18,dullemond18,villenave25}. For each disk model, we choose the wind-driven accretion strength $\alpha_\text{wind}$, so that turbulent and wind-driven accretion together reproduce the prescribed initial total accretion rate, $\dot{M}_\text{acc,init}$.

The paper is organized as follows. In Section \ref{sec:methods}, we summarize the numerical disk model and describe our implementation of an MHD disk wind-driven accretion and mass loss in the \textit{DustPy} code. In Section \ref{sec:results}, we present the resulting dust size distributions, dust-to-gas mass ratios, and potential SI-favorable regions for the parameters probed in our parameter study. In Section \ref{sec:discussion}, we discuss the implications and limitations of our results. In Section \ref{sec:conclusion}, we summarize our conclusions.


\section{Methods} \label{sec:methods}

Our goal is to investigate dust evolution over 1 Myr in magnetized protoplanetary disks that experience turbulent accretion, wind-driven accretion, and wind-driven mass loss. For that, we use the one-dimensional finite volume disk evolution code \textit{DustPy} to model dust coagulation/fragmentation and dust transport, and modify the gas evolution to include the effect of magnetized disk winds following the formalism of \citet{chambers19}.Below we summarize the relevant components of \textit{DustPy} (see \citet{stammler22} for a more detailed overview) and explain how we implement wind-driven accretion and wind-driven mass loss in the code. 

\subsection{Gas evolution}

The gas in our model is treated as a single fluid encoded in the gas surface density ($\Sigma_g$), which evolves under a radial advective-diffusive evolution:

\begin{equation}
    \frac{\partial}{\partial t} \Sigma_g + \frac{1}{r} \frac{\partial}{\partial r} \left (r \Sigma_g v_g \right) = \textcolor{red}{-}\dot{\Sigma}_\text{wind}, 
    \label{eqn:gas_evo}
\end{equation}

\noindent where $v_g$ is the total gas velocity and $\dot{\Sigma}_\text{wind}$ is the outflow due to the MHD wind. $v_g$ is defined as 
 
 \begin{equation}
    v_g =  v_{\mathrm{turb}} + v_w,
    \label{eqn:v_gas}
\end{equation}

\noindent with the first and second terms denoting the turbulence-driven and MHD wind-driven accretion velocities, respectively. Each component of accretion has a corresponding accretion rate: 

\begin{align}
    \dot{M}_\mathrm{acc,turb} &= -2 \pi r_\text{in} \, v_\mathrm{visc} (r_\text{in}) \Sigma_g (r_\text{in}), \\
    \dot{M}_\mathrm{acc,wind} &= -2 \pi r_\text{in} \, v_w (r_\text{in}) \Sigma_g (r_\text{in}),
    \label{eqn:accrate_calcs}
\end{align}

\noindent and the outflow $\dot{\Sigma}_\text{wind}$ has the corresponding mass loss rate 

\begin{equation}
    \dot{M}_\mathrm{wind} = \int_{r_{in}}^{r_{out}} 2 \pi r \dot{\Sigma}_{g,\mathrm{wind}} \ \mathrm{d} r,
    \label{eqn:massloss}
\end{equation}

\noindent where $r_\text{in}$, $r_\text{out}$ are the inner (1 AU) and outer (500 AU) boundaries of our simulations, respectively. The total accretion rate in the disk is the sum of the turbulence and wind-driven components: $\dot{M}_\text{acc,tot} = \dot{M}_\text{acc,turb} + \dot{M}_\text{acc,wind}$.

The turbulence-driven accretion velocity is given by 

\begin{equation}
    v_\mathrm{turb} = -\frac{3}{\Sigma_g \sqrt{r}} \frac{\partial}{\partial r} \left (\Sigma_g \nu \sqrt{r} \right),
    \label{eqn:vel_acc}
\end{equation}

\noindent where $\nu = \alpha_\mathrm{turb} c_s H_p$ \citep{shakura73}. $c_s$ denotes the thermal sound speed of the gas: 

\begin{equation}
    c_s = \sqrt{\frac{k_\text{B} T}{\mu}},
\end{equation}

\noindent where $T$ is the disk temperature at the midplane, $k_B$ is the Boltzmann constant, and $\mu$ is the gas mean molecular weight, taken to be $2.3 m_H$ in all of our runs. $H_P$ denotes the gas scale height: 

\begin{equation}
    H_p = \frac{c_s}{\Omega_K},
    \label{eqn:gas_height}
\end{equation}

\noindent where $\Omega_K$ is the Keplerian orbital frequency. The second term in Equation \ref{eqn:v_gas} is the wind-driven accretion velocity $v_w$.

To quantify the accretion velocity and mass loss rate due to the MHD wind, we use the parametrized model of \citet{chambers19}, where 

\begin{align}
    & v_w = f_w v_0 \left (\frac{T}{T_0} \right)^{1/2} \, , \label{eqn:vwind} \\
    & \dot{\Sigma}_{g,\mathrm{wind}} = \frac{K f_w v_0 \Sigma}{r_0} \left (\frac{r}{r_0} \right)^{-3/2} \, , \label{eqn:outflow}
\end{align}

\noindent in which $f_w$ is the fraction of the total accretion flow due to the wind, $T$ is the local temperature of the disk, $r_0$ is a reference radius, and $T_0$, $v_0$ the temperature and accretion velocity at that reference radius. The equations above show that the mass loss rate in the disk will depend on the fraction of total accretion due to the wind ($f_w$), but the wind-driven accretion velocity $v_w$ will not depend on $K$. The parameter $K$ instead determines the mass loss rate in the wind: $K = 0$ implies that wind-driven mass loss rate is negligible, while larger values of $K$ imply a wind that requires higher mass loss rates to drive the same accretion.

In our models, the reference radius is chosen as $r_0 = 1 \ \mathrm{AU}$ and the reference temperature as $T_0 = 300 \ \mathrm{K}$, from Equation \ref{eqn:T_rad}. By contrast, we change our reference accretion velocity $v_0$ to match the initial accretion rate of our choice. In our Results, we use $\alpha_\text{wind}$ and $\alpha_\text{turb}$ to describe wind-driven and turbulence-driven accretion, respectively. These can be calculated from our model parameters as follows \citep{chambers19}: 

\begin{align}
    \alpha_\text{turb} &= \frac{(1-f_w) r_0 v_0 \Omega_0}{c^2_{s0}} \\
    \alpha_\text{wind} &= \frac{f_w v_0}{c_{s0}} .
\end{align}

\noindent Note that in this formalism, there is no explicit dependence on the magnetic fields; only their effect in driving a portion of the accretion flow is considered and parameterized. Although recent work provided scaling relations between $\alpha_\text{wind}$ and the magnetic field strength in the disk \citep{riols20}, the exact relationship between them remains uncertain. A comparison between numerical solutions of the wind-driven gas surface density evolution of \citet{chambers19} and our own implementation is provided in Appendix \ref{sec:app_a}, for various combinations of $f_w$ and $K$.

In \textit{DustPy}, the temperature profile of the disk is assumed to be that of a passively irradiated disk with a constant irradiation angle of $0.05$: 

\begin{equation}
    T_\mathrm{rad} = \sqrt[4]{\frac{1}{2} \frac{0.05 L_*}{4 \pi r^2 \sigma_{\mathrm{sb}}}},
    \label{eqn:T_rad}
\end{equation}

\noindent where $L_*$ denoting the luminosity of the central star, and $\sigma_{sb}$ is the Stefann-Boltzmann constant. In this work, we assume $L_* = L_\odot$ in all cases. 

Our model numerically solves for the radial evolution of the gas column density $\Sigma_g$ using Equation \ref{eqn:gas_evo}. The vertical distribution of gas is not numerically calculated, and the gas is instead assumed to be in vertical hydrostatic equilibrium, with its scale height given by Equation \ref{eqn:gas_height}. The volume density of gas in the midplane is then 

\begin{equation}
    \rho_\text{g,mid} = \frac{\Sigma_g}{\sqrt{2\pi}H_P}\textcolor{red}{.}
    \label{eqn:}
\end{equation}

The initial gas disk profile is that of the self-similar disk of \citet{lynden-bell74}. We note that this is an analytical solution to the time-independent diffusive equation (Equation \ref{eqn:gas_evo}) and does not necessarily represent physical disks.  

\begin{equation}
   \Sigma_{g,\mathrm{init}} = \frac{M_\mathrm{disk}}{2 \pi \, R_c^2} \left (\frac{r}{R_c} \right)^{-1} \exp{\left [ - \left (\frac{r}{R_c} \right)\right ]}, 
    \label{eqn:init_gas}
\end{equation}

\noindent where $M_\text{disk}$ is the initial mass and $R_c$ is the exponential cut-off radius. 

We use a constant gradient inner boundary and a floor value outer boundary in all of our simulations. A disk with strong wind-driven accretion can be advective across its entire extent, as opposed to a viscously evolving disk which is diffusive beyond $R_c$. This raises the possibility that gas can flow from beyond the outer boundary of the simulation and feed the inner regions of the disk. We do not consider this and assume that there is no material to be accreted beyond the outer boundary of our simulations. Our set-up parameters for the gas disk are provided in the top panel of Table \ref{tab:disk_params}.

\subsection{Dust evolution}

Unlike the single-fluid gas that is evolved in the model, the dust contains multiple species characterized by their mass, which are simultaneously evolved in the disk via coagulation/fragmentation and transport. 

The radial motion of each dust species \textit{i} is evolved using an advection-diffusion equation \citep{clarke88}: 

\begin{equation} \label{eqn:dust_evo}
    \begin{aligned}
    &\frac{\partial \Sigma_{d,i}}{\partial t} + \frac{1}{r} \frac{\partial}{\partial r} \left (r \Sigma_{d,i} v_{d,i} \right) \\ 
    - &\frac{1}{r} \frac{\partial}{\partial r} \left [ r D_i \frac{\partial}{\partial r} \Sigma_g \left ( \frac{\Sigma_{d,i}}{\Sigma_g} \right ) \right ] = S_\mathrm{ext,i},
    \end{aligned}
\end{equation}

\noindent where the subscript \textit{i} denotes the particular dust species whose evolution is governed by the above equation. The second term on the LHS of Equation \ref{eqn:dust_evo} governs dust advection, while the third term governs diffusion. $v_{d,i}$ is the radial velocity of dust species \textit{i}: 

\begin{equation}
    v_{d,i} = (v_g - 2\eta v_K \text{St}_i) \frac{1}{1+\text{St}^2_i}, 
\end{equation}

\noindent where $\text{St}_i$ is the Stokes number, and $\eta$ is related to the radial pressure gradient in the disk as 

\begin{equation}
   \eta =  -\frac{1}{2} \left ( \frac{H_P}{r} \right )^2 \frac{\partial \log P}{\partial \log r}.
    \label{eqn:eta}
\end{equation}

\noindent The value of the Stokes number $\text{St}_i$ depends on whether dust is in the Epstein or Stokes I regime, and is given by

\begin{equation}
    \text{St}_i = \begin{cases}
        &\frac{\pi}{2} \frac{a_i \rho_\mathrm{mon}}{\Sigma_g} \ \text{if} \ a_i < \frac{9}{4}  \lambda_{\text{mfp}} \ (\text{Epstein})\\
        &\frac{2\pi}{9} \frac{a_i^2 \rho_\text{mon}}{\lambda_\text{mfp} \Sigma_g} \ \text{if} \ a_i \ge \frac{9}{4} \lambda_\text{mfp} \ (\text{Stokes I}), 
    \end{cases}
    \label{eqn:stokes}
\end{equation}

where $\lambda_\text{mfp}$ is the mean free path of the gas, 

\begin{equation}
    \lambda_\text{mfp} = \frac{1}{\sqrt{2} \: n \sigma_{\text{H}_2}},
\end{equation}

\noindent with $n$ the midplane gas number density, and we take the material density of dust to be $\rho_\text{mon} = 1.67 \ \mathrm{g/cm^2}$ in all cases.

$D_i$ in Equation \ref{eqn:dust_evo} is the dust diffusivity, given by \citep{youdin2007}:

\begin{equation}
    D_i = \frac{\delta_r c_s^2}{\Omega_\mathrm{K}} \frac{1}{1+\mathrm{St}_i^2}
\end{equation}

\begin{table*}[ht]
\centering
\begin{tabular}{|c|c|c|c|c|c|c|c|c|c|}
\hline
 & \multicolumn{9}{|c|}{List of Simulations} \\ 
 \hline
 $\dot{M}_\text{acc,tot,init}$ & \multicolumn{3}{|c|}{$1\rm{e}-7 \ M_\odot/\text{yr}$} & \multicolumn{3}{|c|}{$1\rm{e}-8 \ M_\odot/\text{yr}$} & \multicolumn{3}{|c|}{$1\rm{e}-9 \ M_\odot/\text{yr}$} \\ 
 \hline
 $\alpha_\text{turb}$ & $1\rm{e}-3$ & $1\rm{e}-4$ & $1\rm{e}-5$ & $1\rm{e}-3$ & $1\rm{e}-4$ & $1\rm{e}-5$ & $1\rm{e}-3$ & $1\rm{e}-4$ & $1\rm{e}-5$ \\ 
 \hline
 $\alpha_\text{wind}$ & $4.62\rm{e}-3$ & $4.66\rm{e}-3$ & $4.66\rm{e}-3$ & $9.32\rm{e}-6$ & $4.29\rm{e}-5$ & $4.62\rm{e}-5$ & \text{\sffamily X}& $9.33\rm{e}-7$ & $4.29\rm{e}-6$ \\
 \hline
 $K = 0.0$ & \checkmark & \checkmark & \checkmark & \checkmark &\checkmark &\checkmark & \text{\sffamily X} & \checkmark & \checkmark \\
 \hline
 $K = 0.2$ & \checkmark & \checkmark & \checkmark & \checkmark &\checkmark &\checkmark & \text{\sffamily X} & \checkmark & \checkmark \\ 
 \hline
$K = 0.4$ & \checkmark & \checkmark & \checkmark & \checkmark &\checkmark &\checkmark & \text{\sffamily X} & \checkmark & \checkmark \\
\hline
$K = 0.6$ & \checkmark & \checkmark & \checkmark & \checkmark &\checkmark &\checkmark & \text{\sffamily X} & \checkmark & \checkmark \\
\hline

\end{tabular}
\caption{A full list of our simulations. We have opted to use the total accretion rate to set the balance between the strength of turbulence and the strength of the wind, and kept our values of $\alpha_\mathrm{turb}$ within observational constraints. After setting the values of $\dot{M}_{\mathrm{acc,tot}}$ and $\alpha_\mathrm{turb}$, we then choose $\alpha_\mathrm{wind}$ such that the combination of the two satisfies our choice of accretion rate. Note that for $10^{-9} \ M_\odot/\mathrm{yr}$), our higher values of $\alpha_\mathrm{turb}$ are no longer valid as they lead to overly high initial accretion rates.}
\label{tab:simulations}
\end{table*}

\noindent where $\delta_r$ governs the strength of dust radial diffusion (similar to $\alpha_\mathrm{turb}$ for the gas), and $\mathrm{St}_i$ is the Stokes number associated with dust species $i$. In our models, $\delta_r$ is set to the same value as $\alpha_\mathrm{turb}$, although we note the relationship between $\delta_r$ and $\alpha_\text{turb}$ may not be direct.

\textit{DustPy} has a dust coagulation/fragmentation scheme built-in for the dust, allowing for the transfer of mass from one species to another at each time-step of the simulation. \textit{DustPy} considers either coagulation or fragmentation/erosion as possible outcomes of dust-dust collisions in the disk, and computes collisional outcomes by calculating relative velocities between dust particles. Several sources of relative velocity are considered: Brownian motion, azimuthal drift, radial drift, and vertical settling. Growth/fragmentation in each radial cell is then calculated using the Smoluchowski equation \citep{smoluchowski16}, and we refer the reader to \citet{stammler22} for further details of the model. 

Like the gas, the vertical distribution of dust species are assumed to be Gaussian, with the dust scale height of each dust species \textit{i} given by \citep{dubrulle95}

\begin{equation}
    H_{d,i} = H_p \sqrt{\frac{\delta_z}{\delta_z + \mathrm{St}_i}},
    \label{eqn:dust_H}
\end{equation}

\noindent and $\delta_z$ is the vertical settling parameter. Like $\delta_r$, $\delta_z$ is set to the same value as $\alpha_\mathrm{turb}$. We again note that this might not necessarily be the case in physical disks. The midplane volume density of each dust species $i$ is given by 

\begin{equation}
    \rho_\text{d,i,mid} = \frac{\Sigma_{d,i}}{\sqrt{2 \pi} H_{d,i}},
    \label{eqn:dust_density}
\end{equation}

\noindent and we compute the midplane dust-to-gas ratio $\epsilon_\text{mid}$ as follows: 

\begin{equation}
    \epsilon_\text{mid} = \frac{\sum_i \rho_{\text{d,i,mid}}}{\rho_\text{g,mid}}.
    \label{eqn:d2g}
\end{equation}

We also compute representative Stokes numbers $\mathrm{St}_\mathrm{rep}$ at each radius in our simulations. We do this by first computing a density-weighted representative size 

\begin{equation}
    a_\mathrm{rep} = \frac{\sum_i a_i \, \rho_{d,i}}{\sum_i \rho_{d,i}}, 
\end{equation}

\noindent where $a_i, \, \rho_{d,i}$ is the size and mass associated with each mass cell $i$ in the simulation, respectively. From here, representative Stokes numbers are calculated using Equation \ref{eqn:stokes}.

Lastly, we compute estimates of dust sizes in the fragmentation and drift limits. The fragmentation limit represents the size beyond which dust will fragment into smaller sizes instead of growing due to dust-dust collisions, and the drift limit represents the size beyond which dust will rapidly drift into the star before having a chance to grow further. Estimates for both limits are given by \citep{birnstiel12}: 

\begin{align}
    a_\mathrm{frag} &= \frac{2}{3 \pi} \frac{\Sigma_g}{\rho_\mathrm{mon} \alpha_\mathrm{turb}} \frac{v^2_\mathrm{frag}}{c_s^2} \label{eqn:a_frag} \\
    a_\mathrm{drift} &= \frac{2}{\pi} \frac{\Sigma_d}{\rho_\mathrm{mon}} \left (\frac{H_p}{r} \right)^2 \left | \frac{\partial \log P}{\partial \log r} \right |^{-1}. \label{eqn:a_drift}
\end{align}

\noindent Using fragmentation and drift size limits, we also calculate the corresponding fragmentation and drift Stokes numbers: $\text{St}_\text{frag}$ and $\text{St}_\text{drift}$ using Equation \ref{eqn:stokes}.

\begin{figure*}[ht]
    \centering
    \includegraphics[width=\linewidth]{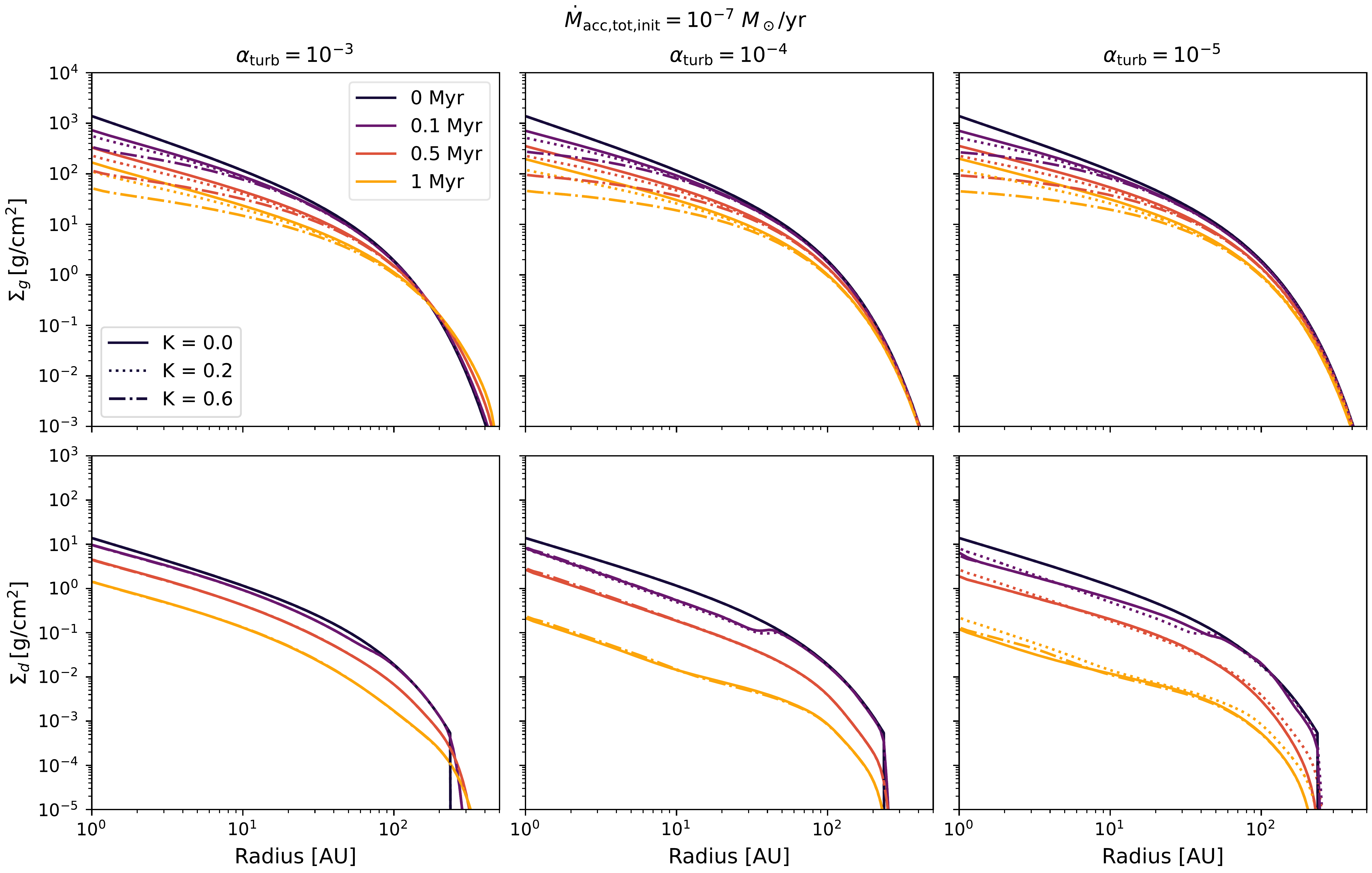}
    \caption{\textit{Top row:} The evolution of gas column densities for various combinations of $K$, $\alpha_\mathrm{turb}$ corresponding values of $\alpha_\text{wind}$ to account for the initial accretion rate, which is $10^{-7} \ M_\odot/$ in all panels of this Figure. We showcase three values of K: $K = 0.0$ (negligible wind mass loss), $K = 0.2$ (moderate wind mass loss), and $K = 0.6$ (high wind mass loss), as a representative sample. Viscous spreading of the disk stops at $\alpha_\text{turb} = 10^{-4}$ in this model as wind-driven accretion begins to dominate in the disk outer regions. Larger wind mass loss rates lead to a reduction in density in the disk inner regions. \textit{Bottom row:} Evolution of corresponding bulk dust densities. The sharp drop-off of density in the outer regions of the disk at t = 0 Myr is because of rapidly drifting particles which are removed in the beginning of the simulation. Importantly, the difference in dust density profiles for different values of K are much less pronounced compared to the gas density evolution.}
    \label{fig:gascoldens}
\end{figure*}

The dust disk is initialized with size profiles according to the interstellar grain distribution of \citet{mathis77}, where $n(a) \propto a^\mathrm{distExp}$, up to a maximum initial size of $a_\mathrm{ini,max}$. We keep the initial disk mass fixed at $M_\mathrm{disk} = 0.05 \: M_\odot$ across all of our runs, and set the initial global dust-to-gas mass ratio to $0.01$. Other fixed parameters used across our simulations, along with their motivations, are provided in Appendix \ref{app:model}. 

A complete list of our simulations, along with associated parameter values ($\dot{M}_\text{acc,tot,init}, \ \alpha_\text{turb}, \ \alpha_\text{wind}, \ K$) is provided in Table \ref{tab:simulations}. Our simulations are set up by fixing the initial accretion rate $\dot{M}_\text{acc,init}$ and turbulence strength $\alpha_\text{turb}$, which then inform us of the $\alpha_\text{wind}$ required to drive said initial accretion. Both $\alpha_\text{turb}$ and $\alpha_\text{wind}$ then remain fixed throughout each of our simulations. 

The parameter $K$ measures the wind-driven mass loss (Equation \ref{eqn:outflow}). We run simulations with a broad set of values for this parameter, ranging from negligible mass loss ($K = 0$; the wind requires negligible outflows to drive accretion) to significant mass loss. Observations of wind mass loss rate to accretion rate ratios ($\dot{M}_\mathrm{wind} / \dot{M}_\mathrm{acc}$) YSOs put it in the range of $0.1-1$ \citep{watson16,alcala17}. In our simulations, the ratio $\dot{M}_\mathrm{wind} / \dot{M}_\mathrm{acc,tot} \sim 0.1$ corresponds to $K = 0.1$, while $\dot{M}_\mathrm{wind} / \dot{M}_\mathrm{acc,tot} \sim 1$ corresponds to $K = 0.6$. We use simulations with $ K = 0.0$ to isolate the influence of wind-driven accretion on dust evolution without including the corresponding mass loss.

\section{Results} \label{sec:results}

We now present the results of our simulations. Our analysis below does not include every single value of the wind mass loss parameter $K$ for the sake of clarity, but showcases a sub-set of disks only. These selected simulation parameters are highlighted in Table \ref{tab:simulations}. 

\begin{figure*}[ht!]
    \centering
    \includegraphics[width=0.85\linewidth]{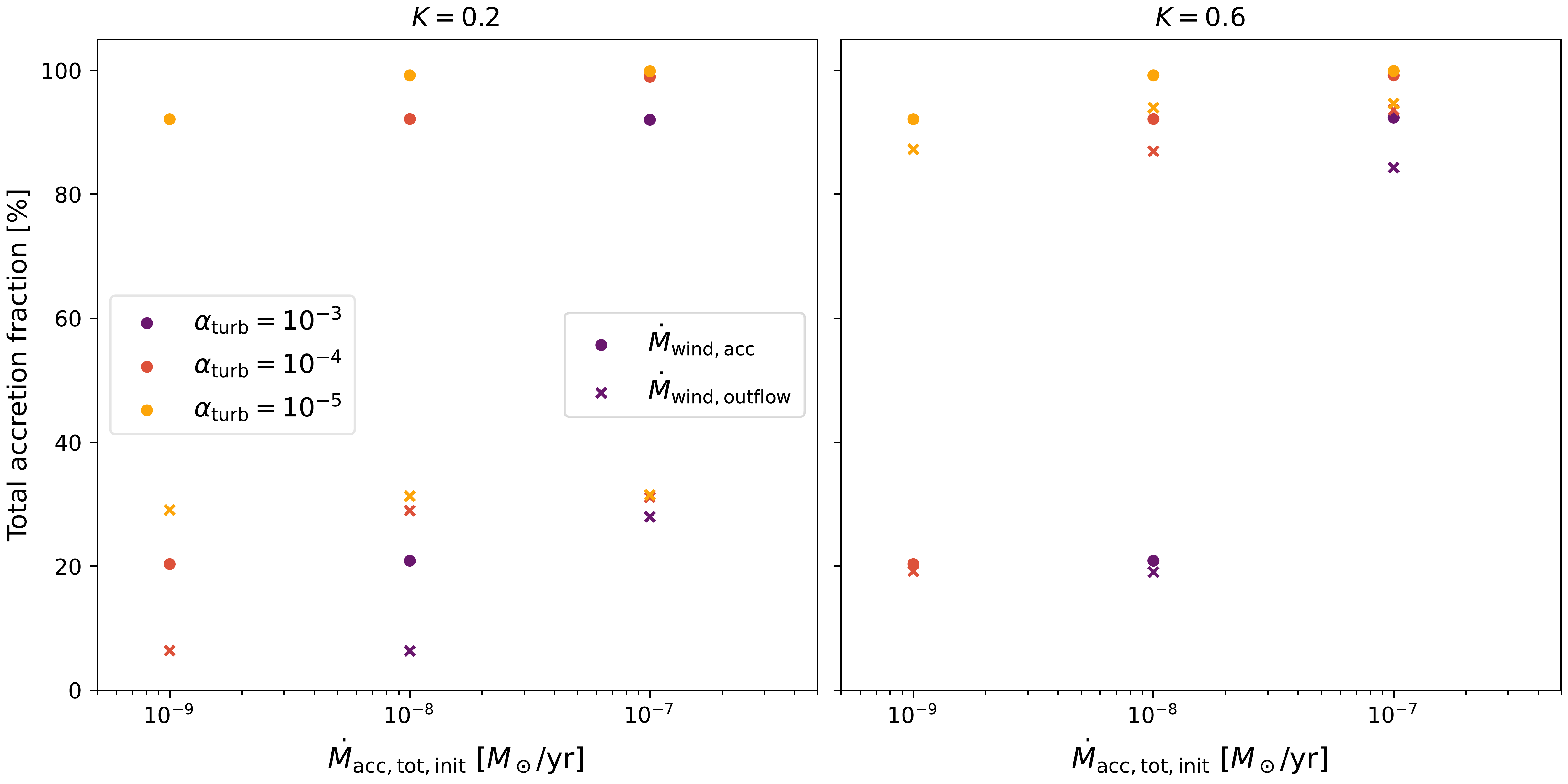}
    \caption{
    Evolution of accretion rates (both turbulent and wind-driven components) and wind-driven mass loss rates. Each row corresponds to different values for the initial total accretion rate, and each column correspond to different values for the wind-driven mass loss parameter $K$. $\dot{M}_\text{wind,acc}/\dot{M}_\text{wind,outflow} \simeq 0.2$ for $ K = 0.2$, whereas $\dot{M}_\text{wind,acc}/\dot{M}_\text{wind,outflow} \simeq 1$ for $ K = 0.6$. Wind-driven accretion rates and mass loss rates as a fraction of the total accretion rates in the disks. The remaining fraction of accretion is driven by turbulence. The two panels correspond to simulations with differing values of the wind-driven mass loss parameter $K$, with $\dot{M}_\mathrm{wind,acc}/\dot{M}_\mathrm{wind,outflow} \simeq 0.2$ for $K = 0.2$ and $\dot{M}_\mathrm{wind,acc}/\dot{M}_\mathrm{wind,outflow} \simeq 1$ for $K = 0.6$.}
    \label{fig:accretion}
\end{figure*}

\subsection{Evolution of Surface Densities And Accretion Rates} 

\begin{figure*}[ht]
    \centering
    \includegraphics[width=\linewidth]{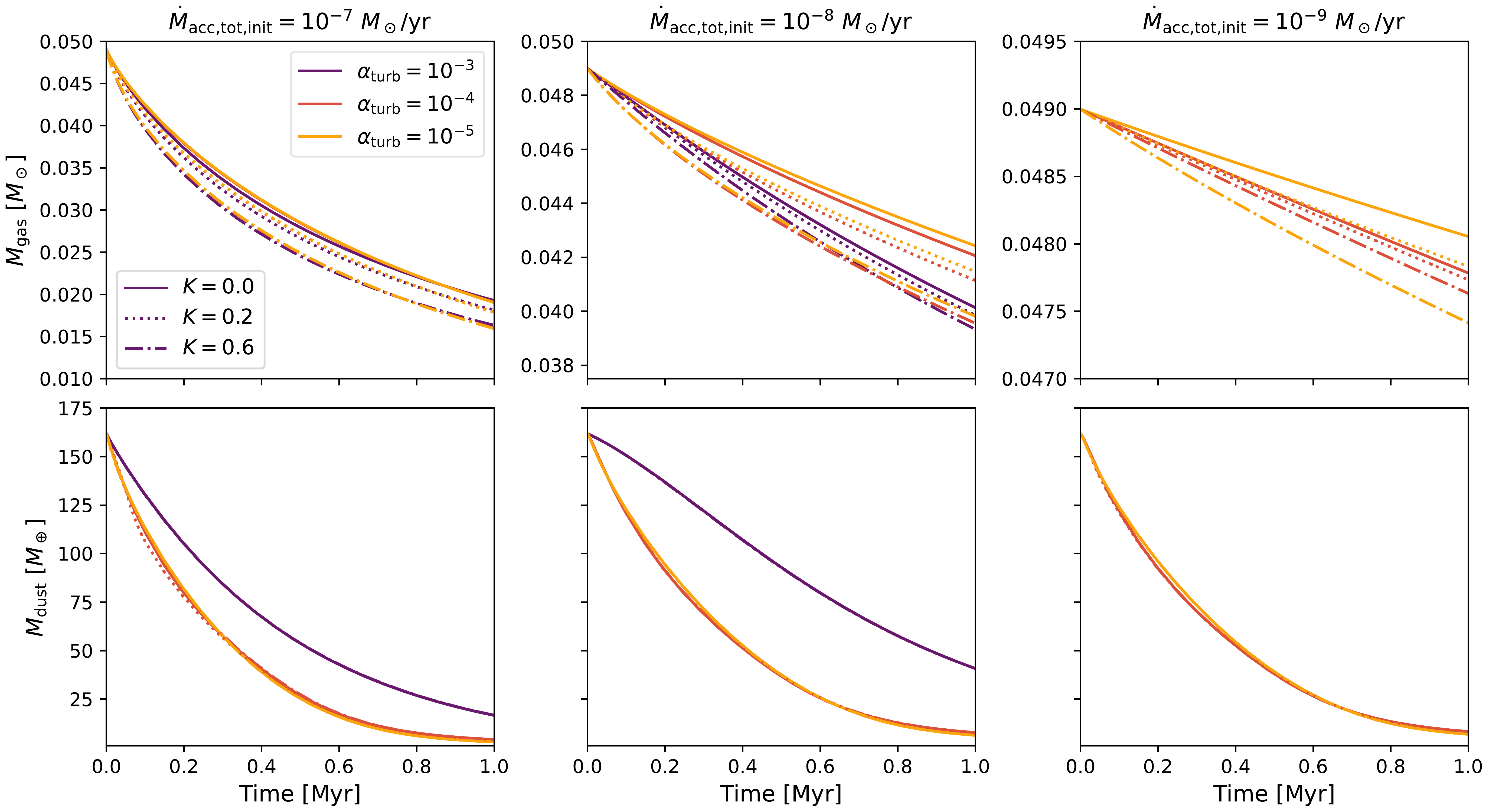}
    \caption{\textit{Top row:} Evolution of total gas mass in our disks across time, where each panel is a different initial total accretion rate. While there is some variance in the mass evolution over time for different values of $\alpha_\text{turb}, \ \alpha_\text{wind}$ and $K$, the behavior is similar across the same $\dot{M}_\text{acc,tot,init}$. \textit{Bottom row:} Evolution of total dust mass in the disks. Unlike the gas mass evolution, which is determined mainly by the initial total accretion and wind mass loss rates, the evolution of dust mass is mainly determined by the strength of turbulence.}
    \label{fig:gasmass}
\end{figure*}

The top row of Figure \ref{fig:gascoldens} shows the evolution of the gas column densities for each choice of $\alpha_\mathrm{turb}$ and $K$ for PPDs with an initial total accretion rate of $\dot{M}_\mathrm{acc,init}=10^{-7} \ M_\odot/\mathrm{yr}$. The shift from disks with diffusive outer regions to advective outer regions occurs at a value of $\alpha_\mathrm{turb}=10^{-4}$, as can be seen in the Figure. The effect of higher mass losses is also apparent, most significantly in the disk inner regions where wind-driven accretion is strongest. Importantly, high values of the wind mass loss parameter $K$ do not affect gas column density profiles as strongly for turbulence-dominated disks (as can be seen in the $\alpha_\mathrm{turb}=10^{-3}$ panel), since the wind mass loss rate scales with the strength of the wind accretion flow in our disk model (Equation \ref{eqn:outflow}).

The bottom row of Figure \ref{fig:gascoldens} shows the evolution of bulk dust column densities, which are calculated by summing dust column densities for each mass bin in the simulation. It can be seen that changing the values of $K$ has a weaker influence on bulk dust densities compared to the gas. This is because our wind mass loss prescription does not include the entrainment of dust in the outflow. Dust with sizes above $1 \ \mu \mathrm{m}$ are not predicted to be entrained by the wind \citep{rodenkirch22,giacalone19}, and since the majority of dust in our simulations quickly grow beyond $1 \ \mu \mathrm{m}$ and expected to have small vertical extent, our exclusion of entrainment does not significantly affect our results. The sharp cut-off of dust in the outer regions of the disk at early times in the simulation is due to the removal of high-St grains, which are expected to rapidly drift into the disk inner regions. The resulting 'wave' of dust creates a large perturbation in the disk and is therefore removed before the beginning of the simulation. Note that the removal of this dust leads to a reduction in the overall dust-to-gas ratio of the disk, which is initially set to $Z = 0.01$. However, the reduction is modest: we find that dust removal from the outer regions leads to an initial dust-to-gas ratio of $Z \simeq 0.099$ across all of our simulations. Initial dust removal therefore does not affect our results, and serves purely to stabilize the early timesteps of our simulations.

In Figure \ref{fig:accretion}, we provide the total accretion fraction of wind-driven accretion and wind-driven outflow, for two choices of the wind mass loss parameter $K = 0.2 \ \mathrm{and} \ K = 0.6.$ The accretion fraction not driven by the wind is driven by the turbulence instead. These fractions remain stable across the runtime of our simulations, even as the total accretion rate decreases over time and the gas column density of the disk drops. Although disks with initial total accretion rates of $10^{-7} \ M_\odot/\mathrm{yr}$ are dominated by wind-driven accretion regardless of the strength of turbulence, lower initial accretion rates can lead to disks which are either dominated by turbulence or by the wind. In addition, we see that wind-driven mass loss rates are at a factor of a few below wind-driven accretion rates for $K = 0.2$, and are about equal to wind-driven accretion for $K = 0.6$.

Figure \ref{fig:gasmass} show the temporal evolution of the total gas mass (top row) and the total dust mass (bottom row). Although the evolution of gas mass in these disks does not show significant changes with weakening turbulence (since wind-driven accretion makes up for weaker turbulence-driven accretion), the main influence on the evolution of dust mass is instead the strength of turbulence. For example, in the case of $\dot{M}_\text{acc,init} = 10^{-7} \ M_\odot/\text{yr}$, the gas mass drops from $0.05 \ M_\odot$ to $\sim 0.02 \ M_\odot$ in $1$ Myr irrespective of the strength of turbulence or mass loss rate. Total dust mass, on the other hand, drops from $160 \ M_\oplus$ to $\sim 20 \ M_\oplus$ for $\alpha_\text{turb} = 10^{-3}$, but drops to $3 \ M_\oplus$ for $\alpha_\text{turb} = 10^{-5}$ by contrast on a 1 Myr timescale. 

\begin{figure*}[ht]
    \centering
    \includegraphics[width=0.63\linewidth]{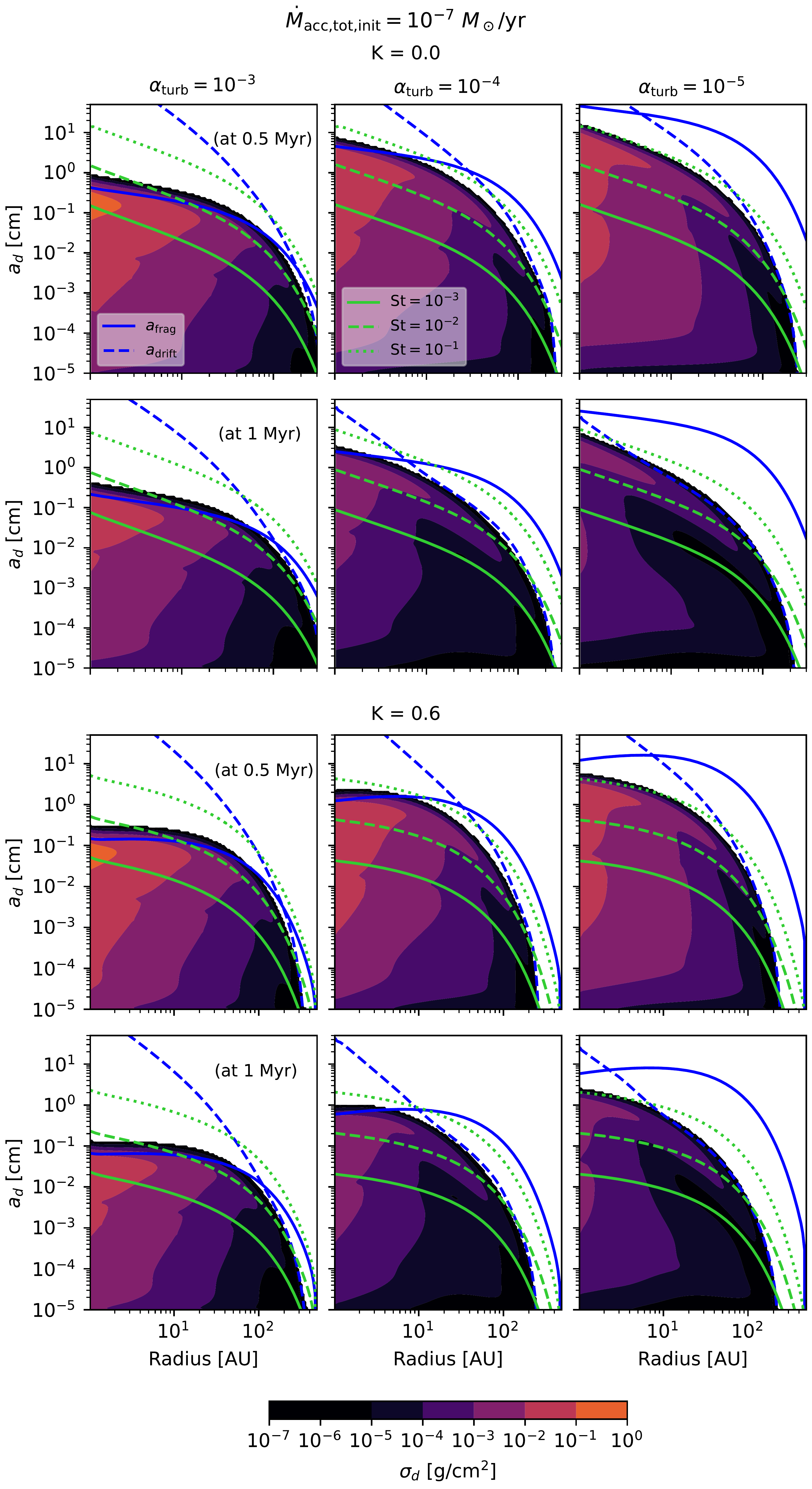}
    \caption{Dust size distribution evolution for $K = 0.0$ and $K = 0.6$ with an initial total accretion rate of $\dot{M}_{\mathrm{acc,init}} = 10^{-7} \ M_\odot/\mathrm{yr}$. The colorbar denotes the surface density corresponding to the dust sizes on the y-axis. Plotted over the distributions are fragmentation and drift limits for growth (blue lines) and various iso-Stokes numbers (green lines). While size distributions show some variation between different values of $K$, the most significant differences are seen with changing $\alpha_\text{turb}$.}
    \label{fig:sizes}
\end{figure*}

\begin{figure*}[t!]
    \centering
    \includegraphics[width=\linewidth]{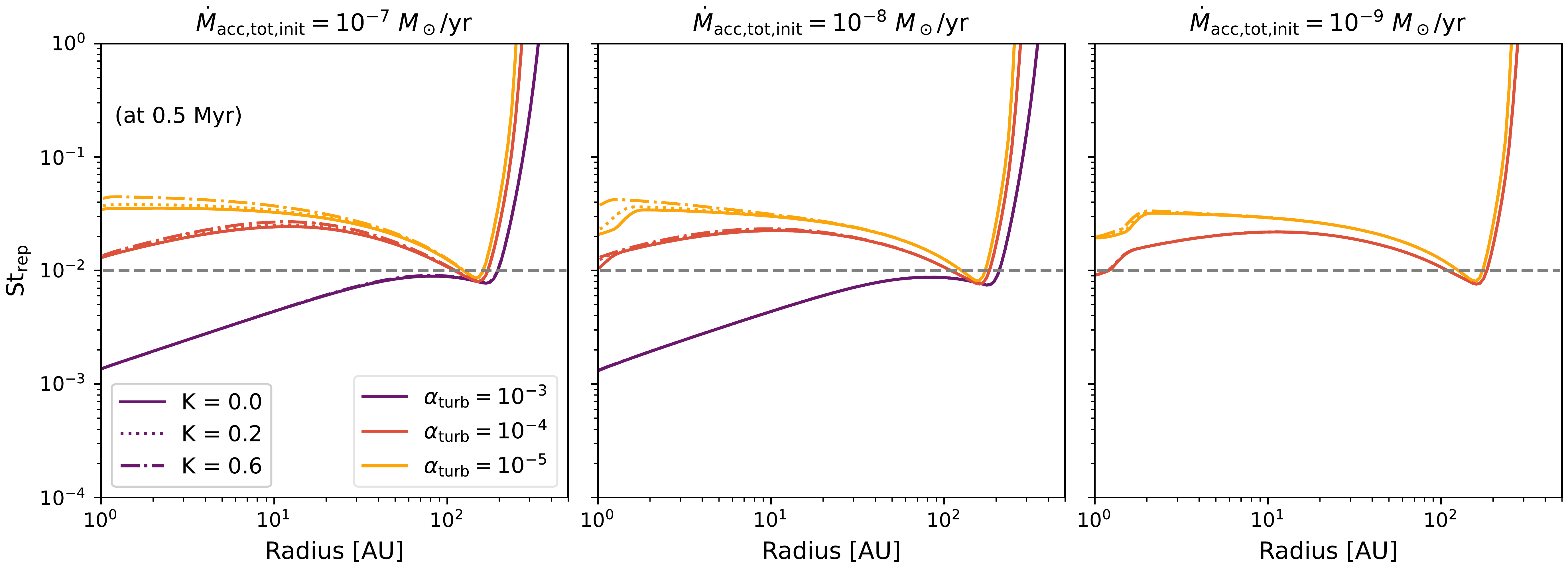}
    \caption{Representative Stokes numbers 0.5 Myr into disk evolution where each panel shows a different initial total accretion rate. A $\mathrm{St}=10^{-2}$ line (gray,dashed) is provided to guide the eye. Stokes numbers are consistently higher when turbulence is weak due to efficient vertical settling and low relative collision velocities between dust particles.}
    \label{fig:stokes}
\end{figure*}

Wind-dominated disks with weak turbulence therefore do not retain their dust as effectively as turbulence-dominated disks due to the higher frequency of collisional interactions from vertical settling into the midplane and the lack of turbulent stirring which would otherwise impede the efficiency of dust growth. In these disks, dust can grow efficiently until aerodynamic drag begins to significantly influence their radial velocity, at which point they are more quickly lost to the central star. Weakly turbulent disks, where accretion is driven primarily by the wind, will therefore be significantly more dust-poor compared to their turbulence-driven counterparts in the absence of a pressure bump to halt radial drift. This also implies that wind-driven disks have a much shorter window of time available to form planets.

\subsection{Evolution of Dust Sizes and Dust-to-gas Ratios}

We now take a closer look at the behavior of dust in our disks, namely their size distributions, representative Stokes numbers, and midplane dust-to-gas ratios. 

Figure \ref{fig:sizes} shows the evolution of dust sizes for an initial total accretion rate of $10^{-7} \ M_\odot/\mathrm{yr}$, for $K = 0.0 \ \mathrm{and} \ K = 0.6$. Each column represents a different balance between turbulence-driven accretion and wind-driven accretion, with turbulence weakening from left to right. The top panel shows results for $K = 0$, representing a wind with negligible mass loss. The bottom panel shows results for $K = 0.6$, representing a wind with high mass loss, with $\dot{M}_\text{wind,acc} \simeq \dot{M}_\text{wind,outflow}$. We also plot in each panel the fragmentation and drift limits for growth (Eqns. \ref{eqn:a_frag} and \ref{eqn:a_drift}, blue curves), in addition to various Stokes numbers ($\text{St} = 10^{-3}, 10^{-2}, 10^{-1}$, green curves). We can see that the strength of turbulence is what mainly determines dust size distributions in the disk, with weakening turbulence (and thus stronger winds to keep to the same accretion rate) corresponding to larger dust sizes. Maximum sizes of dust attained in our disks increase by around two orders of magnitude (from $10^{-1} \ \mathrm{cm}$ to $10^1 \ \mathrm{cm}$) between $\alpha_\mathrm{turb}= 10^{-3}$ and $\alpha_\mathrm{turb}=10^{-5}$, in addition to becoming more concentrated towards the high end of the distribution. Dust sizes begin to shift from being fragmentation-limited to drift-limited at $\alpha_\text{turb} = 10^{-4}$.  

In Figure \ref{fig:stokes} we provide a collection of snapshots of the values of the representative Stokes numbers as a function of disk radius (taken at the  particular time point $0.5$ Myr in the evolution of our disks) for each of our initial total accretion rates. The $\mathrm{St}=10^{-2}$ line is also provided as a reference. Once again, we see the strength of turbulence as the dominant effect in determining the representative Stokes numbers. $\text{St}_\text{rep}$ is largely agnostic to the initial accretion rate and mass loss parameter $K$.

\begin{figure*}[t]
    \centering
    \includegraphics[width=\linewidth]{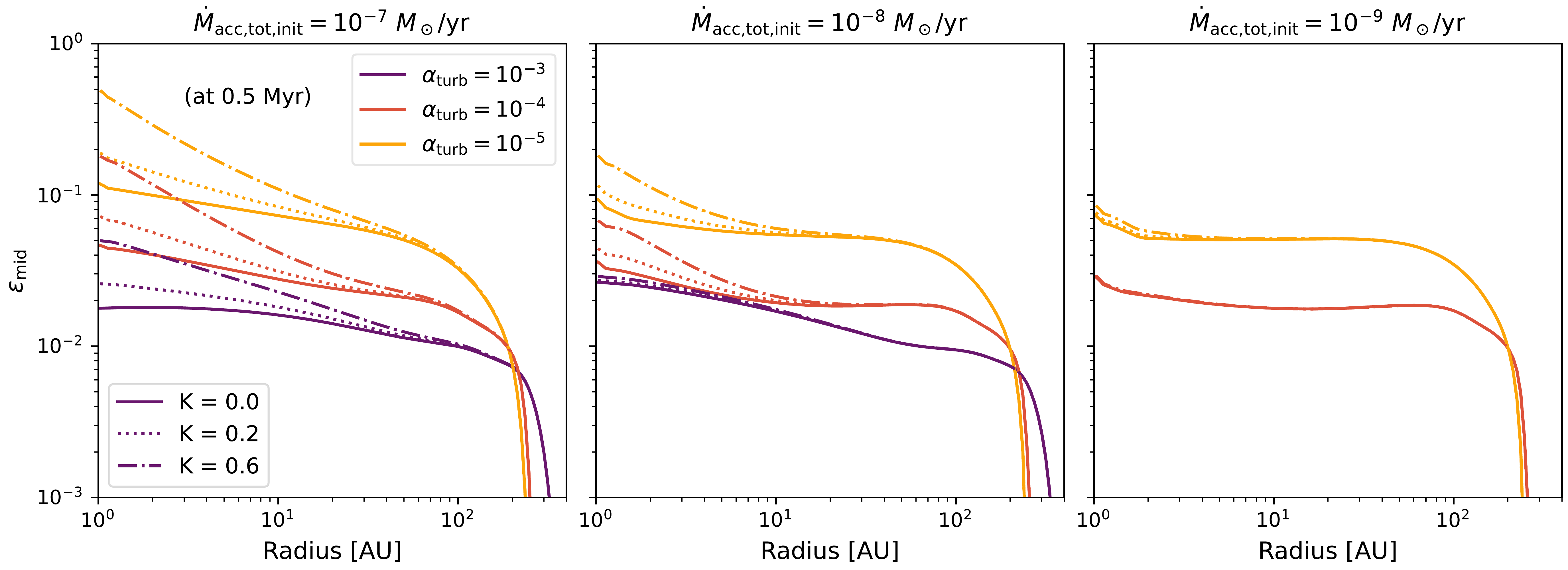}
    \caption{Midplane dust-to-gas ratios 0.5 Myr into disk evolution, where each panel represents a different initial total accretion rate. In addition to steadily increasing dust-to-gas mass ratios with weakening turbulence, we also observe significantly higher dust-to-gas ratios with higher wind-driven mass loss rates.}
    \label{fig:d2g}
\end{figure*}

Finally, we focus on a critical quantity for the onset of the SI and subsequent planetesimal formation: the midplane dust-to-gas ratio. We show these quantities for disks in Figure \ref{fig:d2g} with our different values of the initial total accretion rate. A consistent rise in midplane dust-to-gas ratios mass across the entire radial extent of our disks, especially pronounced near the inner boundary of our simulations, can be seen for decreasing $\alpha_\mathrm{turb}$ in each of our panels (though most pronounced for higher initial total accretion rates in the top row). By contrast, the vertically integrated dust-to-gas ratio ($\Sigma_d/\Sigma_g$) shows the opposite behavior as wind-dominated disks with weak turbulence tend to have lower vertically integrated dust-to-gas ratios when compared to their more turbulent counterparts. This implies that the enhancement in $\epsilon_\mathrm{mid}$ is due to efficient vertical settling when turbulence in the disk is weak. Enhancement of the midplane dust-to-gas ratio with increasing $K$ can also be seen across all of our simulations. This effect is less pronounced with lower initial total accretion rates since the wind mass loss rate associated with high values of $K$ for low accretion rates is also lower, as can be seen in Figure \ref{fig:accretion}. In the most extreme cases of high initial accretion rate and high $K$, midplane dust to gas ratios near 1 AU of our simulations can reach $\sim 0.7$.

\section{Discussion} \label{sec:discussion}

\subsection{Dust Growth And Retention In Weakly Turbulent, MHD Wind-Driven Disks}

Our simulations show that disks with weak turbulence and correspondingly stronger MHD disk wind-driven accretion are significantly more conducive to both the growth and the concentration of dust in the disk midplane. The causes of this are two-fold: the weak turbulence in MHD disk-wind-dominated disks results in more efficient dust settling onto the disk midplane (due to lack of turbulent stirring), as well as increased collision rates  as the dust becomes more concentrated. In addition, the weak turbulence in wind dominated disks lowers collision velocities between dust particles, thus leading to more growth instead of fragmentation.

Figures \ref{fig:sizes} and \ref{fig:d2g} together show the efficient growth and concentration of dust for weakly turbulent, wind-dominated disks. Although both growth and concentration in these disks are promising for planetesimal formation, we also note that such disks are depleted of their dust content when compared to their more turbulent counterparts (Figure \ref{fig:gasmass}, bottom row). Disks are particularly depleted of their dust content in the outer regions of the disk (beyond 10 AU) for $\alpha_\text{turb}=10^{-4}$ and $\alpha_\text{turb}=10^{-5}$. Dust sizes are entirely drift-limited in these regions, implying the rapid growth and subsequent deposition of this dust into the inner disk. By contrast, drift timescales are longer in the inner disk due to fragmentation-limited growth, and dust densities remain higher in these regions compared to the outer disk as a consequence.

A wind-dominated, smooth disk therefore only leaves a few AU in the innermost disk as a possible region of planetesimal formation only under the most extreme circumstances. However, dust is not well-retained even in these regions, as significantly less dust is present at 1 Myr around 1 AU in wind-driven disks compared to turbulence-driven ones. Then, while these disks can both grow and concentrate dust efficiently, the window in which planetesimal formation can take place before dust is lost to the central star is very limited, and formation must take place within a few hundreds of thousands of years.  

\subsection{Prospect of Streaming Instability Onset} \label{subsec:dis_si}

Figure \ref{fig:d2g} shows that midplane dust-to-gas mass ratios can reach values well in excess of the initial global dust-to-gas mass ratio of 0.01, provided that turbulence in the disk remains weak. However, the midplane dust-to-gas mass ratios still remain below $\epsilon_\mathrm{mid} = 1$, which is the value usually taken to be the conservative threshold beyond which SI can occur \citep{johansen07,johansen09}. Follow-up work on SI onset has shown that this ratio can be smaller provided that dust Stokes numbers are high \citep{li21,lim24}. 

Recent local shearing box simulations have shown that threshold midplane dust-to-gas ratios with dust of Stokes numbers above $10^{-2}$ are the most susceptible SI onset \citep{li21,rucska23,lim25}. In particular, \citet{li21} provides a criterion for SI onset as a function of Stokes number:

\begin{equation}
    \log (\epsilon_\mathrm{mid}) \simeq A' (\log \mathrm(St))^2 + B' \, \mathrm{log} (\mathrm{St}) + C',
    \label{eqn:threshold}
\end{equation}

\noindent where 

\begin{align*}
    &A' = 0, \, B' = 0, \, C' = \log{(2.5)} \ \text{if} \ \mathrm{St} < 0.015, \\
&A' = 0.48, \, B' = 0.87, \, C' = -0.11 \ \text{if} \ \mathrm{St} > 0.015
\end{align*}

\begin{figure*}[t]
    \centering
    \includegraphics[width=\linewidth]{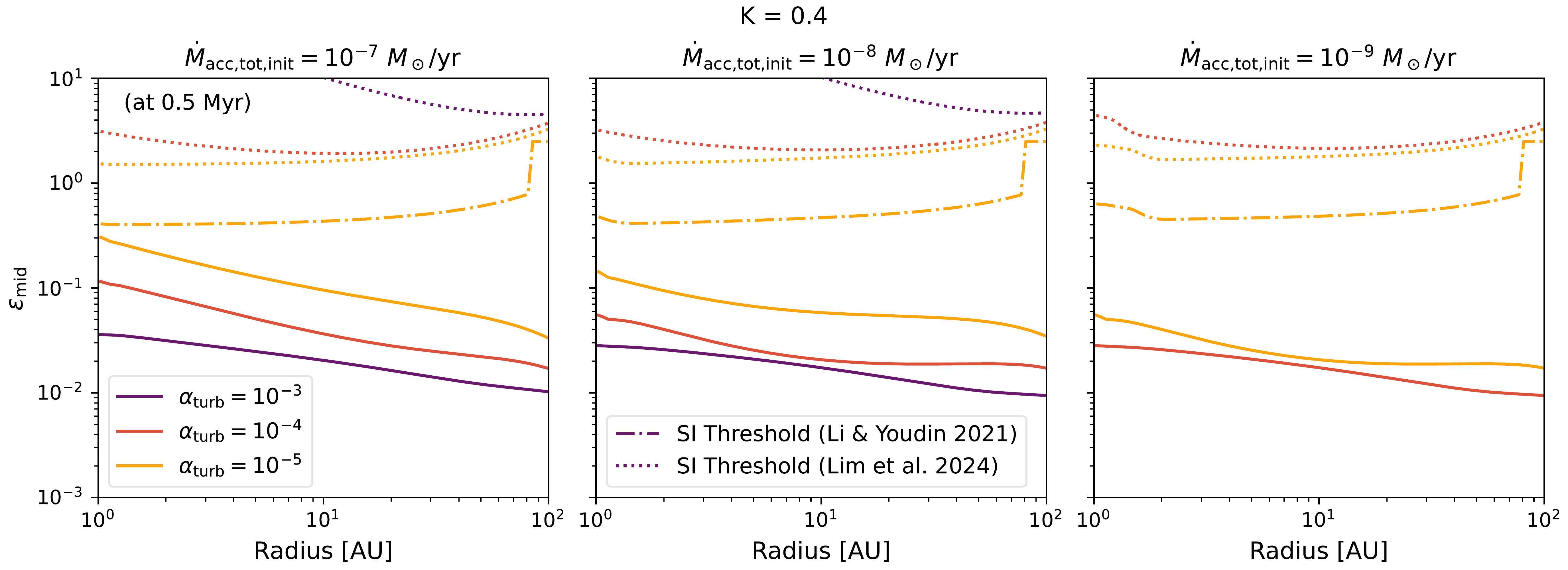}
    \caption{Comparison between the midplane dust-to-gas ratios attained in the simulated disks and SI threshold midplane dust-to-gas ratios provided in \citet{li21} and \citet{lim24}, for all disks with $K = 0.4$. \citet{lim24} thresholds are used for $\alpha_\text{turb} = 10^{-3} \ \text{and} \ 10^{-4}$, while the \citet{li21} threshold is used for $\alpha_\text{turb} = 10^{-5}$. The model of \citet{lim24} includes the turbulence, hence the reason it is prioritized over \citet{li21} thresholds in this Figure. $\alpha_\text{turb} = 10^{-5}$ is not covered in the turbulence range of \citet{lim24}, which is why we show \citet{li21} thresholds for this turbulence strength instead. While the midplane dust-to-gas ratio approaches SI threshold values for $\dot{M}_\text{acc,tot,init} = 10^{-7} \ M_\odot/\text{yr}$ and $\alpha_\text{turb} = 10^{-5}$ near 1 AU, the concentration of dust towards the disk midplane is still not sufficient for planetesimal formation via SI.}
    \label{fig:li_K04}
\end{figure*}

\noindent being the estimate for SI onset in their simulations.

It is important to note that the midplane dust-to-gas ratio thresholds in \citet{li21} do not include the influence of turbulence, and midplane dust-to-gas ratios may need to be greater than unity when the influence of turbulence is considered \citep{lim24}. This threshold dust-to-gas ratio is estimated to be in the same form as Equation \ref{eqn:threshold}, but with coefficients A'=0.42, B'=0.72, and C'=0.37 instead. \citet{lim24} studies the onset of SI using $\alpha_\mathrm{turb}= \{10^{-3}, 10^{-3.5}, 10^{-4}\}$, and do not find a difference with changing values of $\alpha_\mathrm{turb}$. 

We note that both of these works explore disks with  $\Pi = \frac{\eta v_K}{c_s} = 0.05$ that quantifies the radial pressure gradient locally in the disk. We find that all disks we simulate  have $\Pi \sim 0.04$ near the inner boundary, where the midplane dust-to-gas ratio is at its highest, and therefore, most susceptible to planetesimal formation via SI. The radial pressure gradient steepens toward the disk outer regions, where the midplane dust-to-gas ratio is low and SI onset is likely not possible.

\begin{figure*}[t]
    \centering
    \includegraphics[width=\linewidth]{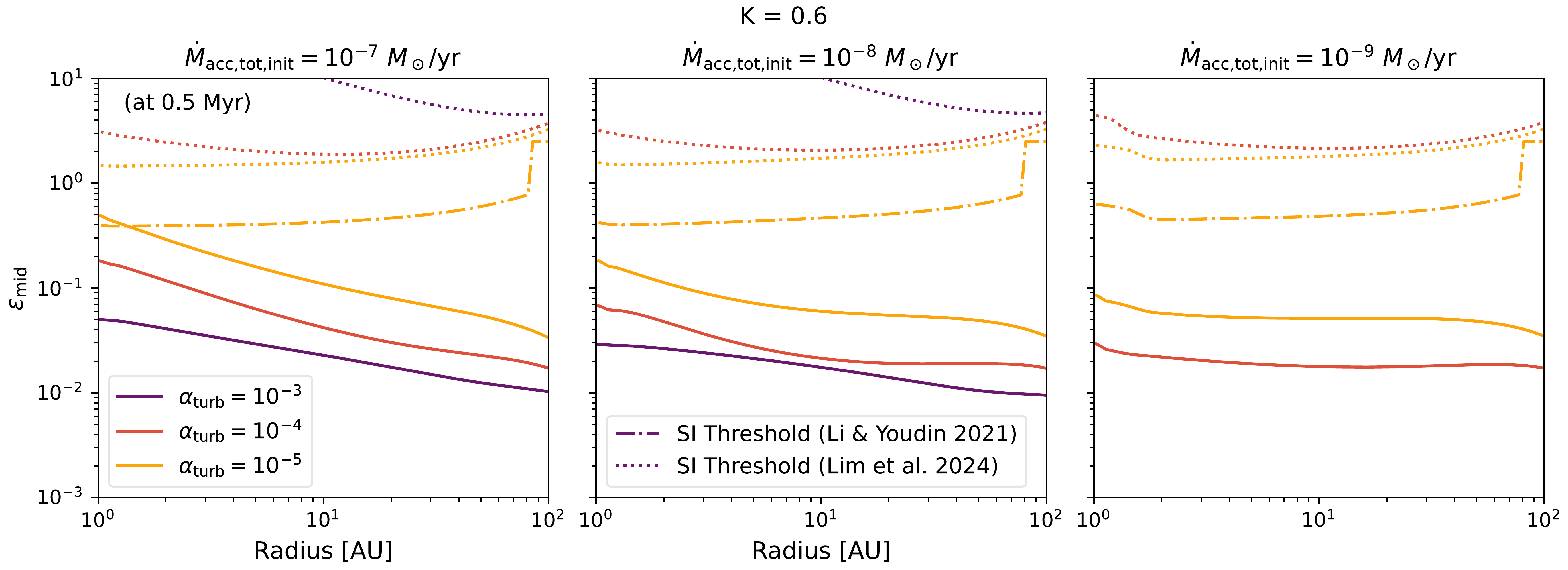}
    \caption{Same as Figure \ref{fig:li_K04}, now showing comparisons between midplane dust-to-gas ratios and \citet{li21} SI thresholds for $K = 0.6$. For these disks, we see that the SI thresholds is crossed in the innermost regions of the disk only for our highest choice of initial accretion rate ($\dot{M}_\mathrm{acc,init} = 10^{-7}$) and weakest turbulence ($\alpha_\text{turb}=10^{-5}$).}
    \label{fig:li_K06}
\end{figure*}

In Figures \ref{fig:li_K04} and \ref{fig:li_K06}, we compare the midplane dust-to-gas ratios of various disks with the SI threshold dust-to-gas ratios of \citet{li21} and \citet{lim24} at $0.5$ Myr. To do this, we calculate a representative Stokes number $\mathrm{St}_\mathrm{rep}$ (using Equation \ref{eqn:stokes}) for each radial grid cell in the simulation, as also shown in Figures \ref{fig:stokes}. We then feed these representative Stokes numbers into Equation \ref{eqn:threshold}, using the set of coefficients A', B' and C' as found in \citet{lim24} for $\alpha_\text{turb}=10^{-3}, 10^{-4}$, and \citet{li21} for $\alpha_\text{turb}=10^{-5}$. This is because \citet{lim24} does not probe turbulence strengths below $\alpha_\text{turb}=10^{-4}$ in their work. 

In each Figure ($K = 0.4$ and $K = 0.6$),  mid-plane dust-to-gas ratios and SI thresholds of disks with different wind-driven mass loss rate parameters are plotted. It can be seen that the SI threshold estimates for $\epsilon_\mathrm{mid}$ of \citet{lim24} are consistently higher than unity regardless of the initial total accretion rate, $\alpha_\mathrm{turb}$ and $K$ that we set for our simulations. None of our simulations have values of $\epsilon_\mathrm{mid}$ that cross unity within our computed radial domain - with our highest value approaching $\sim 0.4$ for $\dot{M}_\mathrm{acc,init}=10^{-7} \ M_\odot/\mathrm{yr}$, $\alpha_\mathrm{turb}=10^{-5}$ and $K = 0.6$. Our only possible candidate for SI formation is therefore a disk with a combination of very weak turbulence and very high mass loss rate. Planetesimal formation by SI therefore appears to be highly unlikely in smooth disks beyond 1 AU, regardless of the disk accretion rate, turbulence strength, strength of wind-driven accretion, and mass loss rate. 

We note that a number of prior studies with similar 1D wind-driven disk models have observed local dust-to-gas ratios in the disk that can overcome the thresholds of SI onset. This can either occur as a result of the creation of a cavity due to the wind-driven outflow \citep{takahashi18,taki21}, or as a result of efficient wind-driven disk cooling which moves the iceline inward, thus allowing for large dust to persist in disk inner regions \citep{kawasaki25}. 

However, we do not observe an inner cavity within our simulation grid (i.e. outward of 1 AU) across the parameter space that we explore. This is likely due to the different wind-driven mass loss prescriptions used in these other models. We do not allow the mass loss rate in the disk to exceed the wind-driven accretion rate in any of our simulations, with our most extreme cases corresponding to $\dot{M}_\mathrm{wind,outflow} \simeq \dot{M}_\mathrm{acc,tot}$. This is supported by the observations which have comparatively lower initial accretion rates ($\dot{M}_\mathrm{acc,tot,init} \simeq 10^{-7} \ M_\odot/\mathrm{yr}$ at its highest) which in turn prevents the rapid clearing of the inner disk by the outflow. Mass loss rates in observed YSOs do not typically exceed accretion rates \citep{watson16,alcala17}. It is therefore unclear how an inner cavity could be created early in the disk lifetime in the absence of a powerful outflow, which may be possibly be provided instead by a magnetothermal wind \citep{bai16,suzuki16}.

It is also important to note that more than one metric can be used to calculate thresholds for SI onset. Local shearing box simulations can cite the vertically integrated dust-to-gas ratio \citep[$Z=\Sigma_d/\Sigma_g$, ][]{li21,rucska23}, the volume density ratio at the midplane \citep[$\epsilon_\mathrm{mid} = \rho_d/\rho_g$, ][]{lim24,lim25}, or both simultaneously. The relationship between $Z$ and $\epsilon_\mathrm{mid}$ will depend on the local vertical distribution of gas and dust, which can change significantly depending on the efficiency of vertical settling. We note, for instance, that while \citet{kawasaki25} observe vertically integrated dust-to-gas ratios above the threshold set out in \citet{li21}, their midplane dust-to-gas ratios never exceed 0.1. By contrast, midplane dust-to-gas ratios in a significant proportion of our simulations exceed 0.1 (Figure \ref{fig:d2g}), but we disregard a majority of these cases as we instead use midplane dust-to-gas ratio thresholds from \citet{lim24} and \citet{li21} as our metric for SI onset. For the time being, more conservative thresholds (i.e. $\epsilon_\mathrm{mid}$ exceeds the canonical threshold of 1) may be more prudent to use in studies such as this that do not directly model the SI mechanism.

Our findings therefore indicate that planetesimal formation via SI is unlikely in laminar disks even as dust growth and concentration are aided by the presence of the disk wind. A pressure bump-generating mechanism is necessary for the creating of dust traps which can become planetesimal nurseries. More physically motivated models of disk evolution are therefore crucial for planetesimal formation in the disk without invoking artifical pressure bumps or initial planetesimal distributions.

\subsection{The Need For More Physically Complete Wind-Driven Disk Models}

In Section \ref{subsec:dis_si}, it is clear that the onset of SI can only occur in the innermost regions of disks that have the rather extreme combination of very weak turbulence ($\alpha_\text{turb}=10^{-5}$) and very high wind-driven mass loss rates ($\dot{M}_\text{wind,outflow} \simeq \dot{M}_\text{wind,acc}$). Planetesimal formation within our parametrized hybrid disk framework is therefore highly unlikely, thus necessitating more complete disk evolution models to capture the beginning of the planet formation process.

Importantly, MHD winds have been shown to create pressure maxima within disks, as has been shown in a number of simulations \citep{suriano17,suriano18,riols20} as well as with linear stability analysis \citep{riols19}. These pressure maxima lead to the pile-up of dust and are ideal locations for planetesimal formation via SI. Yet, these wind-generated pressure maxima cannot form within a parametrized MHD wind framework, and require the dynamic evolution of the magnetic field in addition to the evolution of gas and dust. Non-ideal MHD effects appear to be crucial for the creation of these pressure maxima \citep{riols19}, thus further requiring the inclusion of an ionization chemistry framework \citep{delage23}.

Regardless of the fact that the onset of SI is highly unlikely within the parametrized MHD wind framework, our work still shows that the formation of planetesimals can be more easily achieved if the disk is predominantly wind-driven. Dust concentrations in the midplane are already quite high (Figure \ref{fig:d2g}) if the turbulence is weak, meaning a strong pressure maximum may not be required to push the midplane dust-to-gas ratio beyond SI thresholds. Pressure maxima are also shown to be leaky in both 2D and 3D simulations of dust dynamics \citep{zhu12,huang25} where a combination of turbulent diffusion and vertical dust flow can lead to dust escaping the pressure maxima. Efficient vertical settling and dust growth seen in wind-dominated disks can make dust less susceptible to such leakage, but we note here that dust motions in the vertical direction are not included in our model. 

Finally, we note that MHD winds are not the only mechanism by which pressure maxima in disks can be created. Sublimation, gravitational instabilities and the MRI active-dead zone transition can all lead to pressure maxima as well \citep{saito11,takahashi14,pinilla16}. Regardless of whether the cause is the wind itself or some other mechanism, the MHD wind can be expected to aid in dust concentration and planetesimal formation within pressure maxima.

\subsection{Model Limitations}

This paper presents a first step into the complex problem of magnetic field and disk wind effects on dust evolution and consequences for planet formation.  Its virtue is the focus on the simple but important question of how dust evolves in unstructured disks undergoing a mixture of viscous and MHD wind-driven accretion. We deliberately did not include the dynamic evolution of the magnetic field nor the disk microphysics which are the topics of our subsequent papers. 

One advantage of our approach is that the wind mass loss removes only the gas from the disk and not the dust. Thus our model represents the optimal case for dust concentration leading to possible planetesimal formation in wind-driven disks. This is a simplification, as dust can be entrained in wind \citep{rodenkirch22}, although the maximum size of dust that can be entrained in MHD winds is around $1 \ \mu \text{m}$. Although most of the dust is of sizes greater than 1 micron for many of our disks, the entrainment of sub-micrometer dust in the MHD wind would serve to reduce our dust-to-gas ratios in the midplane as shown in Figure \ref{fig:d2g}.  The basic point of our result is that even under the optimal conditions for exciting SI, the midplane dust-to-gas ratios still fail to reach SI thresholds - even when large amounts of gas are being removed with the outflow.

The main limitation of this first model is clear: it is unable to capture the back reaction of the magnetic field on the disk gas and the dust. This interplay is crucial for the creation of wind-driven substructures seen in higher-dimensional simulations \citep{suriano17,bethune17} and is the focus of our next paper. In addition, the level of ionization in the disk controls the level of non-ideal MHD effects \citep[Ohmic diffusion, ambipolar diffusion, and the Hall effect, ][]{nakano86,wardle99}, which in turn affects both the evolution of the magnetic field and the strength of the MHD wind. The level of ionization in the disk is dictated by disk microphysics, where dust size distributions can significantly impact the ionization chemistry \citep{delage22,delage23}. Both the dynamic magnetic field evolution and the ionization chemistry will be included in our wind-driven disk model in future work. 

Our model also does not include the backreaction of dust on the gas. While dust will not affect gas dynamics if the local dust-to-gas ratio is low, this is no longer the case in regions of high dust concentration. In particular, high local dust-to-gas mass ratios can lead to a reduction in the inflow of gas, thus serving to further enhance dust-to-gas ratios \citep{garate20}. Dust begins to alter the dynamics of gas for a combination of high Stokes numbers, high midplane dust-to-gas ratios and low turbulence, with $\text{St} \: \epsilon_\text{mid}/(\epsilon_\text{mid} + 1) \sim \alpha_\text{turb}$ the threshold beyond which dust backreaction becomes significant \citep{kanagawa17}. For more weakly turbulent disks with $\alpha_\text{turb} \leq 10^{-4}$ (where $\text{St}_\text{rep} > 10^{-2}$) this threshold is crossed, implying that dust backreaction may alter the midplane dust-to-gas ratios we see in the disk. This may serve to counter the aforementioned reduction in dust-to-gas ratio due to the entrainment of dust in MHD wind.

Finally, we assume that the disk is vertically isothermal, passively irradiated, and does not undergo viscous heating. The temperature structure of the disk is highly consequential for our wind model, as the wind-driven accretion velocity $v_w$ depends on the temperature profile. Although the heating of the disk is dominated by stellar radiative heating beyond the heat-transition region (where the rates are comparable), it becomes relevant for the disk inner regions where gas column densities and turbulence-driven accretion rates are higher \citep{lecar06}. However, since the strength viscous heating depends both on the strength of turbulence and the dust's contribution to the disk opacity \citep{bitsch15}, its influence is not significant for disks most conducive to planetesimal formation by SI, which are characterized both by very weak turbulence ($\alpha_\text{turb}=10^{-5}$) and large dust grains. In addition, while relaxing our assumption of a vertically isothermal disk would affect the behavior of wind-driven accretion across our simulations, such an implementation is beyond the scope of this paper.

\section{Conclusion And Future Work} \label{sec:conclusion}

In this work, we investigate whether protoplanetary disks with magnetized winds can generate regions of both enhanced dust-to-gas mass ratios and suitable particle sizes in which planetesimal formation via the streaming instability could occur. We built a parametric hybrid disk evolution model that includes turbulent accretion, wind-driven accretion, and wind-driven mass loss. For that, we use the one-dimensional disk evolution code \textit{DustPy} to model dust coagulation/fragmentation and dust transport, and modify the gas evolution to include the effect of magnetized disk winds. We used it to model the evolution of the disk on a timescale of $1\,$Myr, and carry out a parameter study in which we vary the initial total accretion rate over $\dot{M}_\text{acc,init} = 10^{-7}-10^{-9} \ M_\odot/\text{yr}$, the wind-driven mass loss rate from negligible to heavy ($\dot{M}_\text{wind,outflow} \simeq \dot{M}_\text{wind,accretion}$), and the turbulence strength over $\alpha_\text{turb} = 10^{-5}-10^{-3}$. For each disk model, we choose the wind-driven accretion strength $\alpha_\text{wind}$, so that turbulent and wind-driven accretion together reproduce the prescribed initial total accretion rate, $\dot{M}_\text{acc,init}$. Our findings can be summarized as follows:

\begin{itemize}

    \item The size of solid dust particles in wind-dominated, weakly turbulent disks are significantly larger than turbulence-dominated ones. From $\alpha_\mathrm{turb}=10^{-3}$ to $\alpha_\mathrm{turb}=10^{-5}$, coupled with larger $\alpha_\mathrm{wind}$ to keep the disk to the same initial accretion rate, we observe maximum dust sizes to be enhanced by around two orders of magnitude (from $a_d$ = 0.1 to 10 cm) in the inner regions of the disk. 
    \item Wind-driven disks are more dust-poor compared to turbulence-driven ones. Higher radial drift velocities as a result of efficient vertical settling and growth in disks with weak turbulence leads to disks being drained of their dust content more rapidly. The total gas mass in the disk, on the other hand, evolves similarly regardless of the dominant accretion mechanism in the disk.
    \item Midplane dust-to-gas ratios are enhanced in weakly turbulent wind-driven disks, through a combination of efficient vertical settling of large dust grains and the removal of gas from the disk via wind-driven outflows. With a combination of high initial accretion rate ($\dot{M}_\text{acc,init}=10^{-7} \ M_\odot/\text{yr}$), weak turbulence ($\alpha_\text{turb}=10^{-5}$) and strong outflows ($\dot{M}_\text{wind,outflow} \simeq \dot{M}_\text{wind,acc}$), midplane dust-to-gas mass ratios approach $\sim 0.4$ at around 1 AU. 
    \item While planetesimal formation by the streaming instability may be possible around the inner boundary of our simulations (1 AU), this can only occur in a very narrow range of the parameter space that we explore ($\alpha_\text{turb}=10^{-5}$, $K = 0.6$).
\end{itemize}

Even if wind-driven disks are more conducive both to dust growth and concentration towards the midplane, these disks cannot in general form planetesimals in the outer regions of the protoplanetary disks in the absence of pressure maxima halting radial drift and concentrating solid dust particles. Such pressure bumps may themselves be created due to the influence of the wind on the disk. Self-arrangement of magnetic field lines into regions of higher concentration, thus enhancing both the mass loss and wind-driven accretion flows, have been observed in a number of simulations \citep{bethune17,riols20, suriano18}, along with observations of layered outflows from HL Tau \citep{bacciotti25}. Such regions of enhanced magnetic field concentration can open gaps in the disk, thus leading to pressure maxima and possible locations for the onset of SI, and consequently, planetesimal formation.

Thus, the MHD disk winds may provide an alternative mechanism for the creation of rings and gaps in disks without needing to invoke the existence of a pre-existing planet. The parametrized MHD wind model we employ here cannot create such structures, and requires the inclusion of both dynamic magnetic field evolution and non-ideal MHD effects. Such a model, and pressure bumps created therein to allow for planetesimal formation, will be the focus of our future work. 

\begin{acknowledgments}
We thank Tommy Chi Ho Lau, Mario Flock and Til Birnstiel, and we thank John Chambers for providing us with details regarding their hybrid disk model, which allowed us to compare our implementation to their numerical solutions. AG is supported by McMaster University. REP is supported by a Discovery Grant from the National Science and Engineering Council (NSERC) of Canada. TND is supported by The Royal Society Career Development Fellowship (grant number CDF\textbackslash R1\textbackslash 241766). The computational work in this paper was achieved on the Fir computing cluster (previously Cedar), part of the Digital Research Alliance of Canada computing network, under resource allocation code ycg-982-af.  
\end{acknowledgments}

\begin{contribution}

Olcay Ates Goksu was responsible for modifying the software (DustPy) to include the influence of the MHD wind, running models, generating Figures, conducting analysis of results, and writing the main body of this manuscript. Ralph E. Pudritz was responsible for funding and supervising this work, guiding each step of the project, analysis of results and editing this manuscript. Timmy N. Delage was responsible for suggesting the software used for this work, providing many of the scripts used in the implementation of the hybrid disk model, providing essential guidance on implementation, providing critical feedback and suggestions on results and analysis, writing the Abstract and Introduction sections, and editing the main body of this manuscript.


\end{contribution}

%

\software{DustPy \citep{stammler22}, numpy \citep{harris20}, matplotlib \citep{hunter07}, astropy \citep{astropy22}, scipy \citep{scipy20}.}

\newpage

\appendix

\section{Numerical Comparison} \label{sec:app_a}

In this Section, we provide comparisons between numerical solutions of the hybrid gas disk evolution equations in \citet{chambers19} to our own implementation within \textit{DustPy}. Numerical solutions of \citet{chambers19} can be found in the Appendix of their paper. 

\begin{figure*}[ht]
    \centering
    \includegraphics[width=\textwidth]{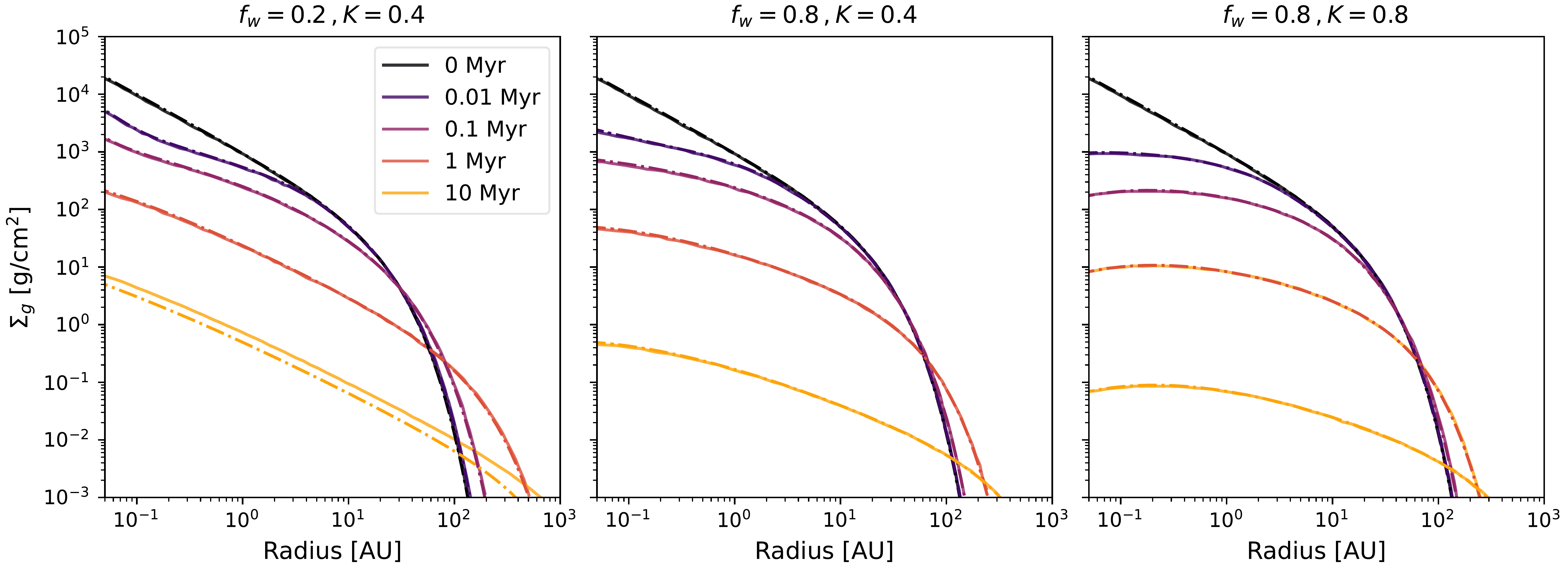}
    \caption{Gas column densities of disks with various combinations of $f_w$ and $K$. Solid lines showcase the evolution of the disk numerically solved by \textit{DustPy}, whereas dot-dashed lines are numerical solutions as shown in \citet{chambers19}, which are very well-matched in all cases.}
    \label{fig:sanity_check}
\end{figure*}

The comparisons can be found in Figure \ref{fig:sanity_check}. Note that the inner boundaries of these simulations are 0.01 AU instead of the 1 AU we use in our simulations, and that the simulations run for 10 Myr instead of 1 Myr. Gas evolution within \textit{DustPy} is not the limiting factor in terms of the computational costs of simulations; it is instead primarily the dust coagulation/fragmentation calculations. The comparisons provided in Figure \ref{fig:sanity_check} were made with the dust evolution in the disk switched off, thus allowing for both a close-in inner boundary and evolving the disk over much longer timescales. 

Three combinations of $f_w$ (which govern the MHD wind-driven accretion flow and is converted to $\alpha_\text{wind}$ in the main text) and $K$ (which govern the mass loss rate) are provided in the Figure to demonstrate the accuracy of our implementation over the relevant parameter ranges. Our implementation and the numerical solutions from \citet{chambers19} are essentially indistinguishable over all radii and across the lifetime of the disk in all cases.

\section{Model Parameters} \label{app:model}

\begin{table*}[ht]
    \centering
    \begin{tabular}{|c|c|c|}
    \hline
    \multicolumn{3}{|c|}{\textit{Gas disk parameters}} \\
      \hline
      \textbf{Parameter}  &  \textbf{Description} & \textbf{Value} \\
      \hline 
      $M_\mathrm{star}$ & Stellar mass & $1 \, M_\odot$ \\
      \hline
       $M_\mathrm{disk}$ & Gas disk mass & $0.05 \, M_\mathrm{star}$ \\
       \hline
       $p$ & Self-similar disk exponent &  -1 \\
       \hline
       $r_c$ & Disk cut-off radius & 50 AU \\
       \hline
       $r_0$ & Reference radius & 1 AU \\
       \hline
       $T_0$ & Temperature at reference radius & 300 K \\
       \hline

       \multicolumn{3}{c}{\vspace{1mm}} \\
       \hline
        \multicolumn{3}{|c|}{\textit{Dust disk parameters}} \\
       \hline
       \textbf{Parameter}  & \textbf{Description} & \textbf{Value} \\
       \hline
        $a_\mathrm{ini, \, max}$ & Maximum dust size in initial distribution & $1 \ \mu m$ \\
       \hline
       distExp & Exponent of initial dust size distribution & -7/2 \\
       \hline
       Z & Initial dust-to-gas ratio & 0.01 \\
       \hline
       $\rho_\mathrm{mon}$ & Dust material density & $1.67 \ \mathrm{g/cm^3}$ \\
       \hline
       $v_\mathrm{frag}$ & Dust fragmentation velocity &$300 \ \mathrm{cm /s }$ \\
       \hline

       \multicolumn{3}{c}{\vspace{1mm}} \\
       \hline
       \multicolumn{3}{|c|}{\textit{Grid properties}} \\
       \hline
       \textbf{Parameter}  & \textbf{Description} & \textbf{Value} \\
       \hline
       Nr & No. of radial grid cells & 128 \\
       \hline 
       $r_\mathrm{min}$ & Inner radial boundary & 1 AU \\
       \hline
       $r_\mathrm{max}$ & Outer radial boundary & 500 AU \\
       \hline 
        $\mathrm{N_{mbpd}}$ & No. of mass bins per mass decade & 7 \\
       \hline
       $m_\mathrm{min}$ & Lower limit of mass grid & $7 \times 10^{-15} \ \mathrm{g}$ \\
       \hline
       $m_\mathrm{max}$ & Upper limit of mass grid & $4.48 \times 10^8 \ \mathrm{g}$ \\
        \hline
    \end{tabular}
    \caption{Set of parameters governing the initial state and evolution of the gas and dust components of our model disk. The top table contains gas disk parameters, while the bottom table contains dust disk parameters. The opacity $\kappa$ is included in the gas parameters, as it only determines the temperature profile of the gas disk.}
    \label{tab:disk_params}
    \end{table*}

We now provide a brief overview of our set-up and our set of simulations. Both physical control parameters kept constant through our simulations and the properties of our simulation grids (both radial and mass) are provided in Table \ref{tab:disk_params}.

The model consists of two grids: a radial grid and a mass grid. The radial grid, spanning 1 AU to 500 AU in the disk, consists of 128 cells spaced logarithmically, and the mass grid spans $7 \times 10^{-15}$ g to $4.48 \times 10^8$ g. Our choice of inner boundary in our radial grid is due to the high computational cost of drawing the inner boundary further in: run times of our simulations go from days to weeks even with small changes to the inner boundary below 1 AU due to the much shorter timesteps required to accurately compute the coagulation/fragmentation of dust at these radii. A smaller choice of inner boundary is therefore too computationally costly for the parameter study we conduct here. By contrast, our choice of the outer radial boundary is motivated by the cut-off radius $R_c = 50$ AU. As the disk is heavily depleted in both gas and dust beyond a few hundreds of AU, the disk outer regions have little influence on dust growth and concentration in the disk inner region, which is the region most relevant to our study.

Our mass grid corresponds to dust sizes ranging from $0.1 \mu \mathrm{m}$ to $4\,$m. A mass grid with a larger upper bound is not required in our simulations as dust will either fragment or be lost to radial drift before reaching such high masses. Our choice of lower boundary limits dust sizes to $0.1 \mu \text{m}$ in keeping with observed dust size distributions in the ISM \citep{mathis77}. The grid itself contains seven cells per decade in mass \citep[a minimum for an accurate coagulation/fragmentation model, ][]{ohtsuki90}, amounting to a total of $162$ cells with our choices of lower and upper bounds.

The fragmentation velocity is uniform across both the radial extent of our disk and across all dust masses, and is set to 3 m/s. The possible range of fragmentation velocities in PPDs is is not well constrained, and could range from $\sim 1 \ \text{m/s}$ to $> 100 \ \text{m/s}$ \citep{wada13,blum10,hasegawa21}, depending on dust properties such as composition and porosity. In addition, comparison between dust continuum emission in the ALMA AGE-PRO sample \citep{zhang25} and simulations of dust evolution are best-matched for a fragmentation velocity of 1 m/s as opposed to their second choice of 10 m/s. We therefore pick $v_\text{frag} = 3 \ \text{m/s}$, though we emphasize once more that fragmentation velocities in physical disks may show large differences depending on size and composition. All of our simulations are initialized with an ISM-like dust size distribution. Note that this assumption is not valid unless significant dust growth has taken place during the core-collapse phase.

The central star has a mass of $1 \ M_\odot$, and we choose the total mass of the disk to be $M_\text{disk} = 0.05 M_*$. This choice for $M_\text{disk}$ is in keeping with the range of $M_\mathrm{disk}/M_\mathrm{star}$ of $1-10 \%$ calculated from rotation curve modeling of the exoALMA sample \citep{longarini25}, however it is important again to point out that, like DSHARP, the exoALMA sample may not be representative of overall PPD populations.

Our disk cut-off radius is located at $50$ AU. Observations of disk radii from dust continuum observations of PPDs are broad in range, ranging from $< 10 \ \mathrm{AU}$ to hundreds of AU \citep{vandermarel21}. Our choice of 50 AU is therefore broadly representative of the radii of the bulk of observed PPD samples.

\section{SI Thresholds at $1$ Myr}

\begin{figure*}[h]
    \centering
    \includegraphics[width=\linewidth]{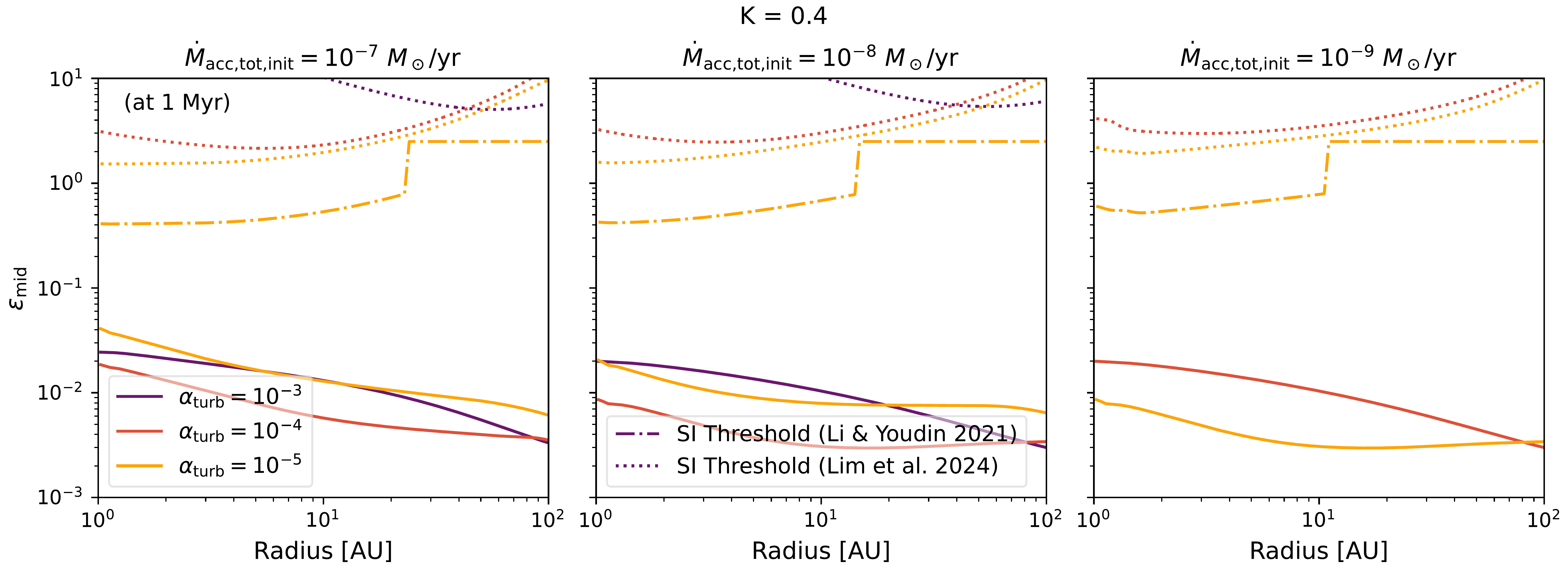}
    \caption{Comparison of midplane dust-to-gas ratios in the disk to SI thresholds of \citet{lim24} and \citet{li21} for all disks with mass loss rate parameter $K = 0.4$ 1 Myr into the disks' evolution. When compared to Figure \ref{fig:li_K04} (which shows 0.5 Myr), midplane dust-to-gas ratios are much lower at this point in the disks' lifetime.}
    \label{fig:li_K04_1myr}
\end{figure*}

\begin{figure*}[h]
    \centering
    \includegraphics[width=\linewidth]{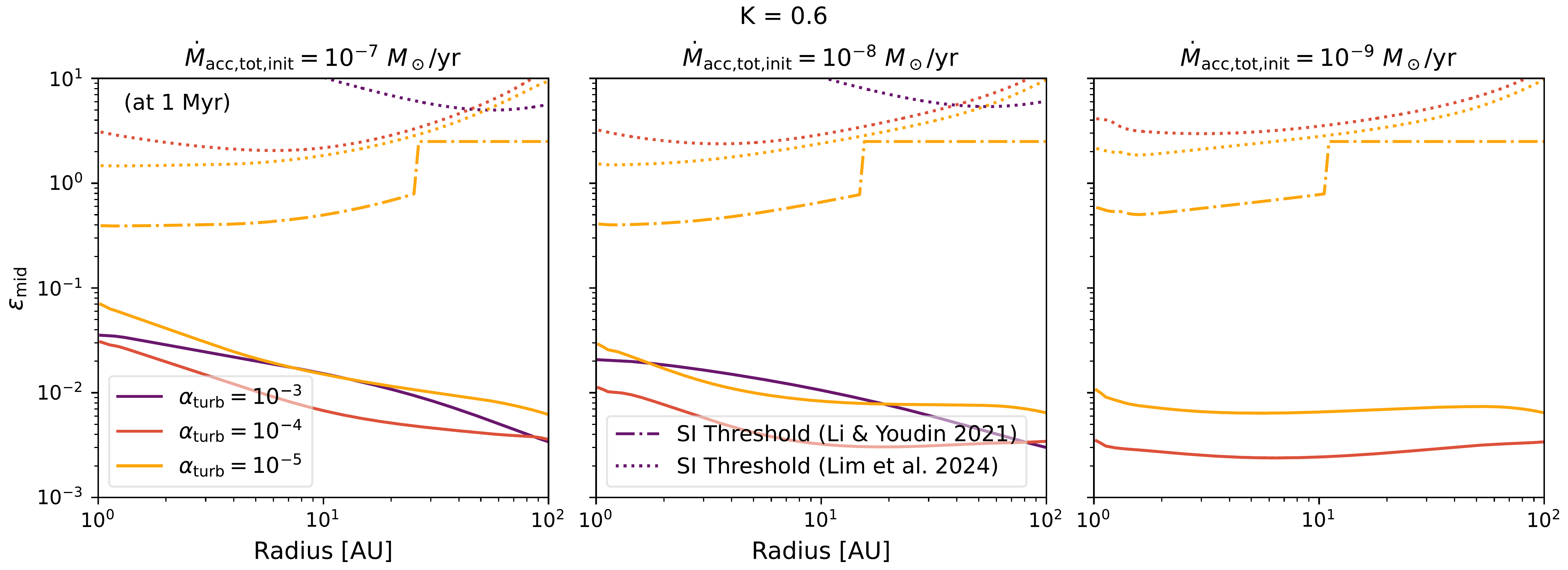}
    \caption{Same as Figure \ref{fig:li_K04_1myr}, this time showing all disks with mass loss rate parameter $K = 0.6$: the highest mass loss we allow in our simulations. The same trend as the $K = 0.4$ is here with the evolution of midplane dust-to-gas ratios. While one particular disk ($\dot{M}_\text{acc,tot,init}=10^{-7} \ M_\odot/\text{yr}$, $\alpha_\text{turb}=10^{-5}$) showed the onset of SI to be possible at 0.5 Myr at 1 AU, midplane dust-to-gas ratios have dropped off significantly by 1 Myr.}
    \label{fig:li_K06_1myr}
\end{figure*}

We now provide comparisons between midplane dust-to-gas ratios in our simulated disks and the SI thresholds of \citet{li21} and \citet{lim24}1 Myr into the disks' evolution, with the same format as Figures \ref{fig:li_K04} and \ref{fig:li_K06}. 

Figures \ref{fig:li_K04_1myr} and \ref{fig:li_K06_1myr} clearly demonstrate that the window for any possible planetesimal formation via SI is brief in the disks that we simulate. Notably, while midplane dust-to-gas ratios are enhanced in weakly turbulent, wind-dominated disks for the first 0.5 Myr, the radial drift of large grains leads to dust depletion in the disk, and subsequently, the midplane dust-to-gas ratios drop off significantly by 1 Myr. 

\section{Dust Behavior With Differing $\delta$}

Here we present a few test simulations wherein we de-couple $\delta$ (which governs the radial, vertical, and turbulent mixing of dust) from the value $\alpha_\mathrm{turb}$. In the simulations presented in the main body of this text, $\delta=\alpha_\mathrm{turb}$ by contrast. 

\begin{figure*}[ht]
    \centering
    \includegraphics[width=0.95\linewidth]{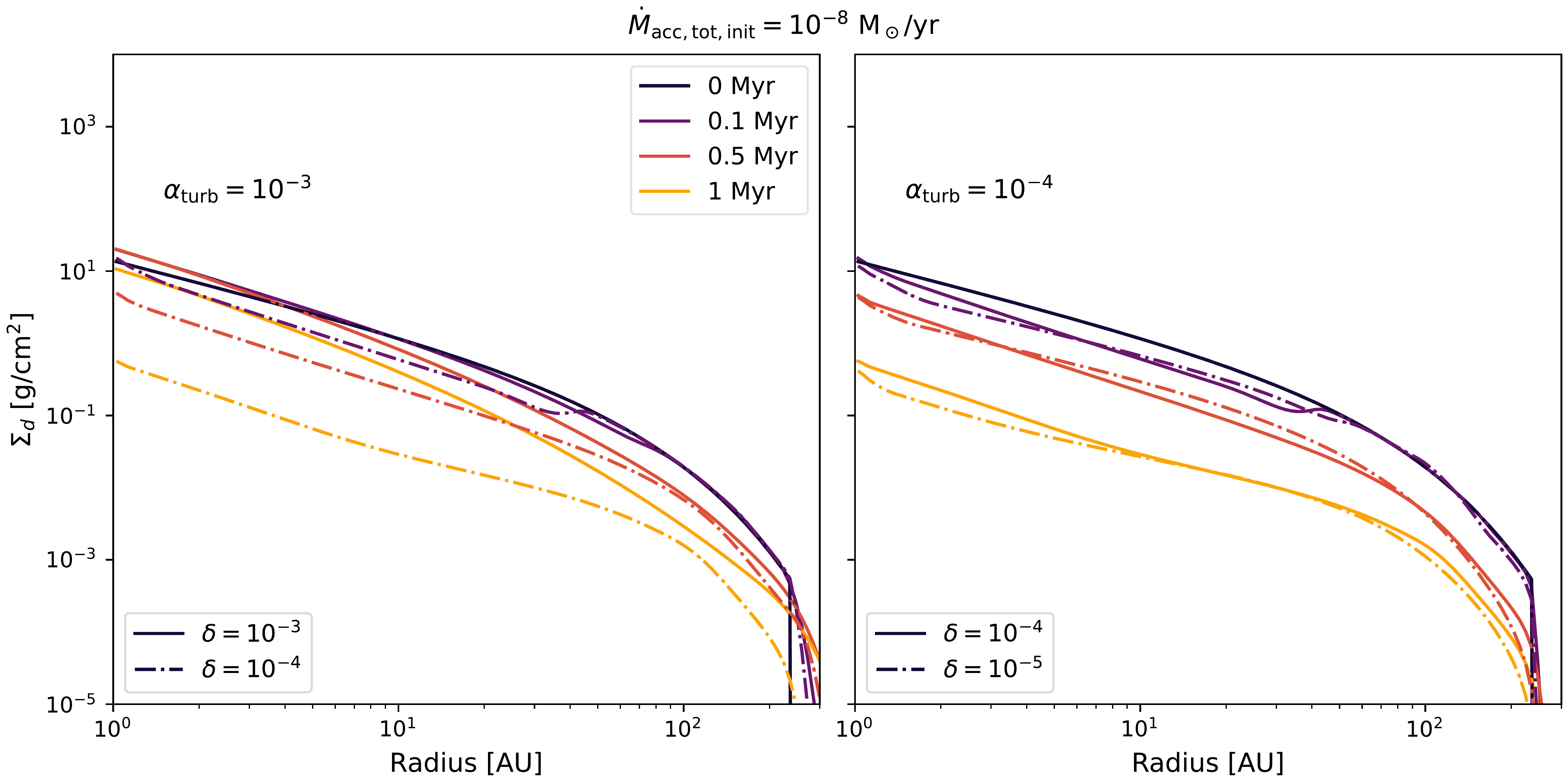}
    \caption{Evolution of dust bulk densities, similar to the bottom row of Figure \ref{fig:gascoldens}. Both panels show disks with $\dot{M}_\mathrm{acc,tot,init}=10^{-8} \ M_\odot/\mathrm{yr}$, with accretion driven primarily by turbulence on the left panel ($\alpha_\mathrm{turb}=10^{-3}$) and by the wind on the right panel $\alpha_\mathrm{turb}=10^{-4}$. The dust $\delta$ is set to be identical to $\alpha_\mathrm{turb}$ in the case of solid lines, and is reduced by a factor of 10 compared to $\alpha_\mathrm{turb}=10^{-4}$ in the case of dot-dashed lines. Whereas the evolution of dust column densities differ significantly for different choices of $\delta$ in the case of a turbulence-dominated disk, only slight differences can be observed when the turbulence is weak and the wind primarily drives the accretion flow.}
    \label{fig:diffdelta_coldens}
\end{figure*}

Figure \ref{fig:diffdelta_coldens} shows bulk dust densities from a number of simulations with identical initial accretion rates ($\dot{M}_\mathrm{acc,tot,init} = 10^{-8} \ M_\odot/\mathrm{yr}$) but with differing values of turbulent-to-wind driven accretion rates and dust $\delta$. We take $\delta=\alpha_\mathrm{turb}$ as the upper limit of $\delta$ in all cases, identical to the approach taken in \citet{pinilla21}. For each disk (primarily turbulence-driven, left panel, and primarily wind-driven, right panel) we compare two cases: one where $\delta=\alpha_\mathrm{turb}$ and another with $\delta=\frac{\alpha_\mathrm{turb}}{10}$. The difference between the two cases is stark in the case of a turbulence-driven disk \citep[as shown in ][]{pinilla21}: the bulk dust density of the disk is noticeably lower when $\delta=10^{-3}$ as opposed to $\delta=10^{-4}$. Low $\delta$ results in more efficient vertical settling, less diffusive dust, and lower relative velocities between dust particles. Each of these factors aid in growth, thus resulting in more dust being lost to the central star due to drift. By contrast, when the disk is primarily wind-driven and the turbulence is weak, a reduction of $\delta$ leads to a bulk dust behavior broadly similar to a disk with $\delta=\alpha_\mathrm{turb}$. A weakly turbulent disk therefore appears agnostic to different choices of $\delta$ as long as the constraint $\alpha_\mathrm{turb} \geq \delta$ is maintained.

\begin{figure}[h]
    \centering
    \includegraphics[width=0.95\linewidth]{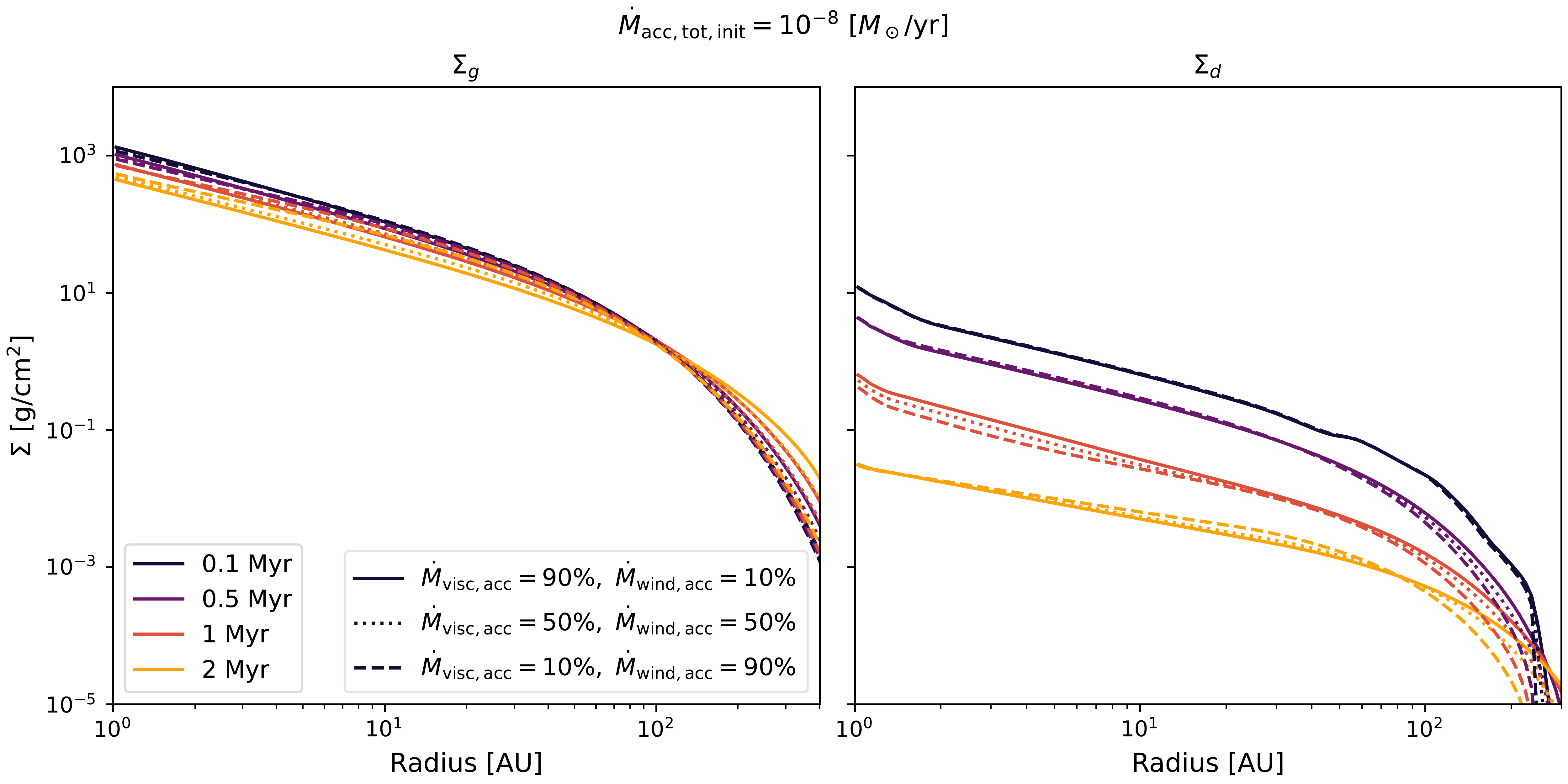}
    \caption{Gas column densities (left panel) and dust column densities (right panel) for disks with the same initial total accretion rate and same dust $\delta$, but with different balances of turbulence-driven accretion and wind-driven accretion. We set a low value for $\delta$ ($10^{-5}$) across each of our runs in order to isolate the influence of changing gas dynamics on the dust profile. Dust behaves broadly similarly across each of the runs.}
    \label{fig:threedisks_coldens}
\end{figure}

We also analyze a number of disks with the same initial total accretion rate and the same dust $\delta$, but with different dominant accretion mechanisms, in Figure \ref{fig:threedisks_coldens}. We considered three cases: a turbulence-driven disk, a wind-driven disk and a hybrid disk (i.e. 50-50 balance between turbulence-driven and wind-driven accretion). There is no mass loss in any of these runs, so $K = 0$. This set-up allows us to isolate the effect of the dominant accretion mechanism on dust evolution. We keep dust $\delta$ at a low value of $10^{-5}$ for all runs. Notably, no significant differences exist between the accretion regimes we analyze. Dust appears to be evolving similarly across all simulations. This is likely due to our choice of $\delta$: efficient growth from vertical settling and low relative velocities with small dust $\delta$ leads to the bulk of dust growing beyond aerodynamic coupling with the gas, and radial drift becomes the dominant mechanism governing dust evolution. 

Figures \ref{fig:diffdelta_coldens} and \ref{fig:threedisks_coldens} therefore indicate that the evolution of the gas disk itself is not very significant for the evolution of dust below $\delta=10^{-4}$. The dust disk exhibits similar behavior even as the physics of the underlying gas disk changes significantly. However, we note here that dust $\delta$ is split into three different components within the model: $\delta_\mathrm{rad}$ (governs radial diffusion), $\delta_\mathrm{vert}$ (governs the degree of vertical settling), and $\delta_\mathrm{turb}$ (governs relative collisional velocities). While each of these parameters can be tweaked individually, we instead opt to keep the same value across all three parameters.


\bibliography{sample701}{}
\bibliographystyle{aasjournalv7}



\end{document}